\documentclass[a4paper,11pt]{article}
\usepackage{jheppub} 
\usepackage{amsmath}
\usepackage{amsfonts}
\usepackage{color}
\usepackage{graphicx}
\usepackage{float}
\usepackage{subfigure}
\graphicspath{{figures/}}
\usepackage{dcolumn}
\usepackage{bm}
\usepackage{extarrows}
\usepackage{amssymb}
\usepackage[numbers,sort&compress]{natbib}
\usepackage{soul}
\usepackage{pifont}
\usepackage{tikz}
\usepackage{sansmath}
\usepackage{mathrsfs}
\usetikzlibrary{shadings,intersections,snakes}
\usepackage{verbatim}
\numberwithin{equation}{section}
\usepackage{tikz}
\usetikzlibrary{calc}

\def\i{{\rm i}}

\def\mD{{\mathcal D}}
\def\re{{\mathrm{Re}}}
\def\im{{\mathrm{Im}}}

\numberwithin{equation}{section} 
\renewcommand{\thesection}{\arabic{section}}

\arxivnumber{2604.02133} 

\title{\bf Effective Field Theory for Superconducting Phase Transitions}

\author[a]{Yanyan Bu}
\author[a]{and Zexin Yang}
\affiliation[a]{\it School of Physics, Harbin Institute of Technology, Harbin 150001, China}

\emailAdd{yybu@hit.edu.cn}
\emailAdd{zexinyang@stu.hit.edu.cn}

\abstract{Employing the Schwinger-Keldysh formalism, we formulate an effective field theory for s-wave superconducting phase transition, where the dynamical variables consist of electromagnetic gauge field and complex scalar order parameter. Symmetry-constrained effective action allows systematic handling of dissipations and fluctuations. In particular, we explore the physical implications of higher-order terms, including those involving additional dynamical fields as well as higher time derivatives, on the real-time dynamics near the superconducting critical point. When appropriately truncated, the effective field theory reproduces the phenomenological Ginzburg-Landau equations. Upon crossing the critical temperature into the low-temperature phase, the electromagnetic gauge symmetry undergoes spontaneous breaking induced by the condensate of the order parameter. Collective excitation analysis reveals that the Higgs mode behaves as an overdamped diffusive mode near the critical point, while the phase fluctuation is absorbed into the gauge field via the Higgs mechanism. For complex kinetic coefficients, we derive the coupled linearized equations for amplitude, phase, and longitudinal gauge fluctuations, and distinguish propagation from damping through their complex poles. The holographic Schwinger-Keldysh construction provides an explicit realization and a nontrivial consistency check of the effective action, and determines its Wilsonian coefficients within a specific strongly coupled model. Holographic results reveal a complex relaxation parameter that indicates oscillatory dynamics characteristic of strongly coupled systems.}

{\footnotetext{Authors are listed in alphabetical order, with both of them to be regarded as both co-first authors and co-corresponding authors.}}

\newpage

\allowdisplaybreaks

\begin{document}
\maketitle
\flushbottom

\section{Introduction}

Superconductivity \cite{tinkham2004introduction} refers to the macroscopic quantum state manifested by certain materials that exhibit zero electrical resistance and perfect diamagnetism when cooled below a specific critical temperature $T_c$. While zero electrical resistance enables persistent electric currents to flow indefinitely without energy dissipation, the Meissner effect, namely the complete expulsion of internal magnetic flux, renders superconductors perfectly diamagnetic.

The Ginzburg-Landau (GL) theory \cite{Ginzburg:1950sr} is a phenomenological framework tailored to describe the superconducting phase transition and the behavior of superconductors near $T_c$. Built on the principles of Landau's theory of second-order phase transitions \cite{Landau:1937obd,Hohenberg:2015jgf}, the GL theory introduces a complex scalar order parameter to characterize the macroscopic quantum signature of the superconducting state. The theory formulates a free energy functional in terms of this order parameter and the electromagnetic vector potential, from which the celebrated GL equations are derived. These equations elegantly capture both the zero-resistance effect and the Meissner effect, and they also predict key phenomena such as vortex formation in type-II superconductors under an external magnetic field \cite{Abrikosov:1956sx}. As a macroscopic framework, the GL theory is independent of the microscopic mechanism underlying electron pairing. It thus provides a universal and effective description applicable across various classes of superconductors, bridging macroscopic experimental observations and microscopic theories.

The microscopic origin of GL theory's order parameter is clarified by the Bardeen-Cooper-Schrieffer (BCS) theory \cite{Bardeen:1957kj,Bardeen:1957mv}, which is the foundational microscopic framework for conventional low-temperature superconductors. The BCS theory probes quantum behavior of electrons in crystalline lattices, revealing that superconductivity arises from phonon-mediated Cooper pairs \cite{Cooper:1956zz}, namely bound states of two electrons with opposite momenta and spins. Below $T_c$, these pairs condense into a coherent quantum state, forming an electronic excitation gap that suppresses scattering by lattice imperfections and impurities and yields zero electrical resistance. Though primarily applicable to conventional low-temperature superconductors, the BCS theory remains a cornerstone of condensed matter physics, establishing a paradigm for studying collective quantum phenomena in many-body systems \cite{Anderson2011BCS,Bruus2004Many}. The BCS theory even inspires the development of subsequent theoretical frameworks for unconventional high-temperature superconductors \cite{Sigrist:1991zz,Wang_2016,Wen_2022,Chen:2022tbp}.

The theoretical underpinnings of superconductivity, specifically the GL theory and the BCS theory, have significantly inspired the development of core concepts and foundational principles in particle physics. In particular, the Meissner effect and the Higgs mechanism are analogous phenomena showing how the concept of spontaneous symmetry breaking, where a field (Higgs field or Cooper pair condensate) acquires a non-zero vacuum expectation value, causes gauge bosons to acquire masses. Thus, the core of superconductivity is the spontaneous breaking of electromagnetic gauge symmetry. The non-Abelian analog of electromagnetic superconductivity, color superconductivity \cite{Alford:2007xm}, was proposed to emerge in the low-temperature and high-density regime of QCD matter.

As a phenomenological framework, the original GL theory can be extended in a number of aspects. Incorporating higher-order corrections in the free energy functional extends the GL theory's validity to lower temperatures and captures finer physical effects \cite{Neumann1966,Neumann1966_2}. Introducing multiple order parameters with distinct characteristic length-scales is crucial for understanding complex materials such as multi-gap superconductors and explaining complex vortex structures \cite{Shanenko2011,Vagov2012,Vagov2012a}. To account for the temporal evolution of superconductors, the original ``static'' GL equations are generalized to the time-dependent Ginzburg-Landau (TDGL) equations \cite{Schmid1966}, which was subsequently justified by Gorkov and Eliashberg \cite{Gorkov_Eliashberg1968} using the BCS theory. Further extensions and refinements of the basic TDGL theory remain active areas of research \cite{Kopnin2001,Harbick2025}. In some extensions, the coefficient of the time-dependent term for the order parameter in the basic TDGL equations is treated as complex, with the imaginary part describing non-dissipative effects such as the propagation of collective modes \cite{Kopnin2001,Larkin2005}. An alternative approach formulates a gauge-invariant, Lorentz-covariant TDGL theory for non-stationary superconductors \cite{Grigorishin:2021gil}, incorporating second-order time derivatives of the order parameter to describe wave-like dynamics instead of pure diffusive relaxation. The TDGL equations and their various extensions prove highly valuable for predicting the dynamical behavior and non-equilibrium properties of superconducting systems \cite{BishopVanHorn2023}. Some key achievements include characterizing fundamental collective oscillations (Higgs and Goldstone modes), forced oscillatory responses to an external gauge field, and dissipative effects inherent to non-equilibrium processes.

Nevertheless, beyond extensions rooted in microscopic theories \cite{Gorkov1959}, all the remaining generalizations elaborated above are inherently phenomenological and thus incomplete. This incompleteness stems from the fact that the GL framework is phenomenological in nature and thus admits a variety of generalizations, each incorporating distinct corrections and approximations.

In this work, we seek to tackle this question using a more robust framework, namely the Schwinger-Keldysh (SK) effective field theory (EFT), which was originally formulated for dissipative fluids \cite{Haehl:2013hoa,Haehl:2015foa,Haehl:2015uoc,Crossley:2015evo,
Glorioso:2017fpd,Haehl:2018lcu,Liu:2018kfw}. In the context of critical superfluidity, a universal TDGL theory was formulated in \cite{Kapustin:2022iih,Donos:2023ibv,Bu:2024oyz,Donos:2025jxb}, wherein stochastic terms are incorporated in a natural manner. Furthermore, this framework \cite{Kapustin:2022iih} enables the systematic inclusion of effects arising from nonuniform temperature and thermal conductivity, as well as explicit or spontaneous breaking of time-reversal symmetry. We aim to extend these prior studies to the realm of superconductivity and construct an EFT for s-wave superconducting phase transition. A key distinction from previous works \cite{Kapustin:2022iih,Donos:2023ibv,Bu:2024oyz,Donos:2025jxb} is that the ``background'' electromagnetic field, which facilitates the construction of superfluid EFTs, is elevated to a dynamical one. 

This work contributes to the broader program of developing a theoretical framework for dynamical electromagnetic fields in general media \cite{Landau1984Electrodynamics}; via the SK EFT formalism, recent studies along this direction can be found in \cite{Vardhan:2022wxz,Landry:2022nog,Das:2022fho,Salcedo:2024nex,Kaplanek:2025moq,Frangi:2025ykb,Yoshimura:2026vil}. Herein, we primarily build on the conceptual idea of \cite{Landry:2022nog} and adopt the superfluid EFT developed in \cite{Bu:2024oyz} as our starting point. Then, we promote the background gauge field to be a dynamical one by integrating over them in the path integral. This is tantamount to gauging the global U(1) symmetry of the initial superfluid system, thereby yielding a corresponding superconducting system\footnote{It is noteworthy that the dynamics of photons in media can be alternatively investigated within the framework of higher-form symmetries \cite{Gaiotto:2014kfa,Grozdanov:2016tdf,Glorioso:2018kcp}, where the photon is regarded as the Nambu-Goldstone boson stemming from spontaneously broken 1-form symmetry \cite{Yoshimura:2026vil}.}.

The rest of this work will be organized as follows. In Section \ref{EFT_action}, we present the SK effective action for an s-wave superconducting system slightly above the critical temperature $T_c$. In Section \ref{Stochastic_Equation_eft}, we demonstrate that our superconducting EFT, once appropriately truncated, reproduces the phenomenological TDGL equations \cite{kamenev_2023}. We demonstrate that the Gorkov-Eliashberg anomalous term \cite{kamenev_2023} emerges naturally within the framework of our EFT. In Section \ref{collective_modes}, we study the dynamics in the low-temperature phase, in which the electromagnetic gauge symmetry undergoes spontaneous breaking induced by the condensate of the order parameter. Collective excitation analysis reveals that the Higgs mode behaves as an overdamped diffusive mode near the critical point, while the phase fluctuation is absorbed into the gauge field via the Higgs mechanism. In Section \ref{lesson_holo_model}, we construct an explicit holographic realization of the EFT and perform a quantitative determination of its Wilsonian coefficients within a specific holographic superconductor. Moreover, holographic results uncover a complex relaxation parameter indicative of oscillatory dynamics, a hallmark of strongly coupled systems. In Section \ref{sec:summary}, we summarize the present study and outline future directions.

\section{Effective action for superconducting phase transition}\label{EFT_action}

In this section, we detail the construction of the effective action for describing s-wave superconducting phase transitions.

The physical scenario can be succinctly abstracted as an electromagnetic field $A_\mu$ within a given superconducting material at finite temperature $T = 1/\beta \sim T_c$. The superconducting phase transition is characterized by a charged scalar field $\mathcal O$, whose non-zero vacuum expectation value induces the onset of superconductivity. We are interested in the behavior of the electromagnetic field at distance and time scales much larger than the characteristic scale $l_{mfp}$. For this purpose, we can imagine integrating out all matter fields to obtain an effective action for $A_\mu$ and $\mathcal O$. To ease the construction, we will take two simplifications: 1) ignore the motion of the medium and temperature fluctuations; 2) assume the temperature of the superconducting material is slightly above the critical one, say $T\gtrsim T_c$.

To address real-time dynamics, we formulate an SK EFT for the superconducting phase transition using the framework developed in \cite{Crossley:2015evo,Glorioso:2017fpd,Liu:2018kfw}. Following the conceptual idea of \cite{Landry:2022nog}, we first imagine that the U(1) symmetry is global and consider spontaneous breaking of the global U(1) symmetry, which corresponds to a superfluid phase transition. Ultimately, we will gauge the U(1) symmetry of the superfluid system, a procedure that yields the superconducting phase transition of our initial interest.

We start by identifying relevant hydrodynamic variables and their parametrization for constructing the superfluid EFT, as developed in \cite{Bu:2024oyz}. To ensure the global U(1) symmetry of the superfluid system, we introduce an external gauge field $A_\mu$ and promote global U(1) symmetry into a local one \cite{Crossley:2015evo}:
\begin{align}
A_{s\mu} \to A_{s\mu} + \partial_\mu \lambda_s(x),  \qquad \mathcal O_s \to e^{\i q \lambda_s(x)} \mathcal O_s,  \qquad s=1~{\rm or}~2, \label{local_U1_extend}
\end{align}
where the variables are doubled in the sense of SK formalism and the index $s$ denotes the two branches of the SK closed time path. Following \cite{Crossley:2015evo}, the hydrodynamical variables for the conserved U(1) current are two neutral scalars $\phi_1, \phi_2$ (i.e., diffusive fields) arising from the promotion of gauge transformation parameters $\lambda_1, \lambda_2$ into dynamical variables. To construct the superfluid EFT, it is convenient to introduce the following building blocks
\begin{align}
B_{s\mu} \equiv A_{s\mu} + \partial_{\mu}\phi_{s}, \qquad \Delta_s\equiv e^{\mathrm{i}q\phi_s} \mathcal O_s, \qquad \Delta_s^*\equiv e^{-\mathrm{i}q\phi_s} \mathcal O_s^*, \label{eq:invariant_vars}
\end{align}
which are invariant under the following local U(1) transformation, namely an extended version of \eqref{local_U1_extend},
\begin{align}
A_{s\mu} \to A_{s\mu} + \partial_\mu \lambda_s(x), \qquad \phi_{s} \to \phi_{s} - \lambda_s, \qquad \mathcal O_s \to e^{\i q\lambda_s(x)} \mathcal O_s
\end{align}
In the Keldysh basis, the $r$- and $a$-variables are defined as:
\begin{equation}
X_{r}\equiv\frac{1}{2}\left(X_{1}+X_2\right), \qquad X_{a}\equiv X_{1}-X_2
\end{equation}
where $X$ collectively denotes the variables $A_\mu$, $\phi$, $\mathcal O$, $B_\mu$ and $\Delta$. The partition function of the superfluid system is a path integral over the diffusive fields $\phi_{r,a}$ and order parameter fields $\mathcal O_{r,a}$
\begin{align}
Z_{sf}[A_{r\mu}, m_r; A_{a\mu}, m_a] = \int [D \phi_r] [D \phi_a] [D \mathcal O_r] [D \mathcal O_a] e^{\i S_{sf}[B_{r\mu}, \Delta_r; B_{a\mu}, \Delta_a]} \label{Z_sf}
\end{align}
where $A_{r\mu}, A_{a\mu}$ represent external sources for the conserved U(1) current and $m_r, m_a$ the sources for the order parameter; and $S_{sf}$ is the desired superfluid EFT action.

The effective action $S_{sf}$ satisfy a list of conditions \cite{Crossley:2015evo,Glorioso:2017fpd,Liu:2018kfw,Bu:2024oyz}:

\textbf{(1) The constraints implied by the unitarity of the time evolution}
\begin{align}
    &\text{Normalization condition:} \quad  S_{sf}[B_{r\mu}, \Delta_r; B_{a\mu} = \Delta_a =0]=0, \nonumber\\
    &\text{Z$_2$ reflection symmetry:} \quad \left(S_{sf}[B_{r\mu}, \Delta_r; B_{a\mu}, \Delta_a]\right)^{*}=-S_{sf}[B_{r\mu}, \Delta_r; -B_{a\mu}, -\Delta_a], \nonumber \\
    &\text{Non-negativity of imaginary part:} \quad \mathrm{Im}\left(S_{sf}\right)\geq0, \label{contraint_unitary}
\end{align}

\textbf{(2) Spatial rotational symmetry:} The action $S_{sf}$ is invariant under spatial rotations, treating the temporal and spatial components of fields distinctly due to the finite temperature medium.

\textbf{(3) Global $U(1)$ symmetry:} Since we are interested in the high-temperature ($T \gtrsim T_c$) phase, the global $U(1)$ symmetry of the superfluid system is not broken spontaneously. So, the SK effective action $S_{sf}$ is invariant under a diagonal global U(1) transformation. This will be automatically satisfied if the action $S_{sf}$ is constructed using the gauge-invariant quantities introduced in \eqref{eq:invariant_vars}.

\textbf{(4) Chemical shift symmetry:} In the normal phase (high-temperature regime), the effective action $S_{sf}$ should be invariant under a time-independent diagonal shift of the diffusive fields:
\begin{align}
\phi_s \to \phi_s + \sigma(\vec x), \qquad {\rm others~unchanged}. \label{chemical_shift}
\end{align}
This symmetry ensures that the effective theory correctly describes charge diffusion in the normal state \cite{Crossley:2015evo}. To reveal the constraint set by the chemical shift symmetry \eqref{chemical_shift}, we explicitly write the gauge-invariant quantities of \eqref{eq:invariant_vars} in the Keldysh basis
\begin{align}
& B_{r\mu} \equiv \frac{1}{2}(B_{1\mu} + B_{2\mu})= A_{r\mu} + \partial_\mu \phi_r, \qquad B_{a\mu} \equiv B_{1\mu} - B_{2\mu}= A_{a\mu} + \partial_\mu \phi_a, \nonumber \\
& \Delta_r \equiv \frac{1}{2}(\Delta_1 + \Delta_2) \approx e^{\i q\phi_r} \mathcal O_r, \qquad \qquad \quad \Delta_a \equiv \Delta_1 - \Delta_2 \approx e^{\i q \phi_r} \left( \mathcal O_a + \i \phi_a \mathcal O_r \right) \label{gauge_invariant}
\end{align}
where in the expressions for $\Delta_r$ and $\Delta_a$, terms beyond linear order in $a$-variables are neglected, corresponding to the semiclassical limit of the EFT in which only thermal fluctuations are captured. Under the chemical shift transformation \eqref{chemical_shift}, the gauge-invariant quantities in \eqref{gauge_invariant} transform as:
\begin{align}
B_{ri} \to B_{ri} + \partial_i \sigma(\vec x), \qquad \Delta_{r,a} \to e^{\i q\sigma(\vec x)} \Delta_{r,a}, \qquad B_{r0} \to B_{r0}, \qquad B_{a\mu} \to B_{a\mu}.
\end{align}
which motivates to introduce the following gauge-covariant derivative of the order parameter:
\begin{align}
\mathcal D_i \Delta_{r,a}\equiv (\partial_i - \i qB_{ri}) \Delta_{r,a}  \label{gauge-covariant_Di}
\end{align}
which under \eqref{chemical_shift} transforms in the same way as $\Delta_{r,a}$:
\begin{align}
\mathcal D_i \Delta_{r,a} \to e^{\i q\sigma(\vec x)} \mathcal D_i \Delta_{r,a}
\end{align}
Then, the chemical shift symmetry \eqref{chemical_shift} places strict constraints on how the spatial derivatives of $B_{ri}$ and $\Delta_{r,a}$ may arise in the effective action: for $B_{ri}$, these spatial derivatives must enter via the gauge field strength $F_{rij}=\partial_i B_{rj} - \partial_j B_{ri}$, whereas for $\Delta_{r,a}$ they must appear through the gauge covariant derivatives $\mathcal D_i \Delta_{r,a}$.

The chemical shift symmetry imposes no constraints on the appearance of time derivatives of all fields or spatial derivatives of the remaining fields.

\textbf{(5) Dynamical Kubo-Martin-Schwinger (KMS) symmetry:} This is the crucial symmetry that enforces the fluctuation-dissipation theorem (FDT) at the full non-linear level. In the classical statistical limit, this symmetry can be stated as
\begin{equation}
S_{sf}[B_{r\mu},\Delta_{r},\Delta_r^*;B_{a\mu},\Delta_{a},\Delta_a^*] = S_{sf}[\tilde {B}_{r\mu}, \tilde{\Delta}_{r}, \tilde {\Delta}_{r}^*; \tilde{B}_{a\mu}, \tilde{\Delta}_{a}, \tilde{\Delta}_{a}^*]\label{SKMSeq}
\end{equation}
where the KMS transformations of each quantities are \cite{Glorioso:2016gsa,Donos:2023ibv,Bu:2024oyz}:
\begin{align}
&\tilde{B}_{r\mu}(-x)=B_{r\mu}(x), \qquad \tilde{B}_{a\mu}(-x)=\left[B_{a\mu}(x) +\mathrm{i}\beta\partial_{0}B_{r\mu}(x)\right], \nonumber\\
&\tilde{\Delta}_{r}(-x)=\Delta_{r}^{*}(x), \qquad \tilde{\Delta}_{a}(-x) = \left[\Delta_{a}^{*}(x) +\i\beta\partial_{0}\Delta_{r}^{*}(x)\right] \nonumber\\
&\tilde{\Delta}_{r}^*(-x)=\Delta_{r}(x), \qquad \tilde{\Delta}_{a}^*(-x) = \left[\Delta_{a}(x) +\i\beta\partial_{0}\Delta_{r}(x)\right]
\end{align}
where $\beta $ is the inverse temperature.  For s-wave superconductivity, the order parameter is a complex scalar, which remains unchanged under $\mathcal{PT}$ transformation but undergoes complex conjugation under the KMS transformation.

\textbf{(6) Onsager relations:} This requirement stems from the symmetry properties of the retarded (or advanced) correlation functions under a change of the ordering of operators \cite{Crossley:2015evo}. In simple cases, Onsager relations are automatically satisfied when KMS symmetry is imposed, though this is not universally valid.

Now we gauge the global U(1) symmetry of the initial superfluid system. This corresponds to promoting the external gauge fields $A_\mu$'s to be dynamical by integrating over them in the path integral \eqref{Z_sf}
\begin{align}
Z_{sc} & = \int [D A_{r\mu}] [D A_{a\mu}] [D \phi_r] [D \phi_a] [D \mathcal O_r] [D \mathcal O_a] e^{\i S_{sf}[B_{r\mu}, \Delta_r; B_{a\mu}, \Delta_a]+ \i S_{\rm Maxwell}} \nonumber \\
& = \mathcal N \int [D B_{r\mu}] [D B_{a\mu}] [D \Delta_r] [D \Delta_a] e^{\i S_{sc}[B_{r\mu}, \Delta_r; B_{a\mu}, \Delta_a]}  \label{Z_sc}
\end{align}
Here, the integrals over the diffusive fields $\phi_r, \phi_a$ drop out through redefinitions of fields, resulting in an infinite constant $\mathcal N$. The superconductor EFT action $S_{sc}$ is identified as
\begin{align}
S_{sc} = S_{sf} + S_{\rm Maxwell}
\end{align}
Since $A_\mu$ is now treated as a dynamical field, we have added its kinetic term
\begin{align}
S_{\rm Maxwell} = \int d^4x \mathcal L_{\rm Maxwell} = - \frac{1}{2} \int d^4x F_{a\mu\nu} F_r^{\mu\nu}
\end{align}
where $F_{r\mu\nu} = \partial_\mu A_{r\nu} - \partial_\nu A_{r\mu} = \partial_\mu B_{r\nu} - \partial_\nu B_{r\mu}$ and similarly for $F_{a\mu\nu}$. Intuitively, $B_{\mu}$ can be taken as the electromagnetic gauge potential in the medium.

The effective action $S_{sc}$ will be organized via a systematic expansion in the number of fields and the number of spacetime derivatives of the fields. The former expansion is motivated by the assumption that the system is near the critical point, such that the order parameter can be treated as small. The second expansion is tied to the fact that we work within the hydrodynamic limit, such that the fields exhibit slow variation over large spatiotemporal scales compared to the characteristic length scale $l_{mfp}$. In addition, the expansion will be truncated up to quadratic order in fluctuations (i.e., $a$-fields) and second order in spacetime derivatives.

For convenience, the effective Lagrangian density, defined by $S_{sc} = \int d^4x \mathcal L_{sc}$, is split as
\begin{equation}
    \mathcal{L}_{\text{sc}} = \mathcal{L}_0 + \mathcal{L}_1. \label{Lscorigin}
\end{equation}
The leading part $\mathcal L_0$ is
\begin{align}
\mathcal{L}_0 = & a_0 B_{a0} B_{r0} + a_1 B_{ai} \partial_0 B_{ri} - \i\beta^{-1} a_1 B_{ai}^2 -\frac{1}{2}F_{a\mu\nu}F_r^{\mu\nu} + b_0 \Delta_a^* \Delta_r + b_0 \Delta_a \Delta_r^* + b_1 \Delta_a^* \partial_0 \Delta_r \nonumber \\
&  + b_1^* \Delta_a \partial_0 \Delta_r^* -2 \i \beta^{-1} \mathrm{Re}(b_1) \Delta_a^*\Delta_a + b_3 \left(\mD_i\Delta_a\right)^* \left(\mD_i\Delta_r\right)+ b_3 \left(\mD_i\Delta_a\right) \left(\mD_i\Delta_r\right)^*  \nonumber \\
&-\mathrm{i} b_3 B_{ai} \left(\mD_i\Delta_r\right)^* \Delta_r + \mathrm{i} b_3 B_{ai} \left(\mD_i\Delta_r\right) \Delta_r^* + c_1 \Delta_a \Delta_r^* \Delta_r^* \Delta_r + c_1 \Delta_a^* \Delta_r \Delta_r^* \Delta_r . \label{L0}
\end{align}

A possible term of the form $B_{a0}(B_{a0} +\i \beta \partial_0 B_{r0})$ has been eliminated by redefinitions of the diffusive fields $\phi_r$ and $\phi_a$. This corresponds to choosing a specific hydrodynamic frame.

From the second condition of \eqref{contraint_unitary}, the coefficients $a_0, a_1, b_0, b_3, c_1$ are real, whereas $b_1$ is allowed to be complex. The last condition in \eqref{contraint_unitary} additionally requires that
\begin{align}
a_1 < 0, \qquad {\rm Re}(b_1) < 0 \label{ImS>0}
\end{align}
Near the critical point, we have $b_0 = \# (T_c -T)$ with a positive coefficient $\#>0$.

If $A_{r\mu}$ is treated as an external source, the dynamical equations for $\phi_r$ and $\mathcal O_r$ derived from $\mathcal L_0$ are equivalent to the stochastic equations of Model F for a critical superfluid \cite{RevModPhys.49.435,Kapustin:2022iih,Donos:2023ibv}, with further details provided in Appendix \ref{superfluid_dynamics}. On the other hand, when $A_{r\mu}$ is regarded as a dynamical field, $\mathcal L_0$ yields precisely the basic stochastic TDGL equations for critical superconductivity, as elaborated in Section \ref{Stochastic_Equation_eft}.

We turn to the remaining part $\mathcal L_1$
\begin{align}
\mathcal{L}_1= & a_{2}B_{a0}B_{r0}B_{r0} + b_2\Delta_a^* \partial_0^2\Delta_r + b_2^* \Delta_a \partial_0^2\Delta_r^* + 2 \beta^{-1} {\rm Im}(b_2) \Delta_a^* \partial_0 \Delta_a + c_0 B_{a0} \Delta_r^* \Delta_r \nonumber \\
& + c_0 B_{r0} \Delta_a^* \Delta_r + c_0 B_{r0} \Delta_a \Delta_r^*   + c_2 B_{r0}^2 \Delta_a^* \Delta_r + c_2 B_{r0}^2 \Delta_a \Delta_r^* + 2 c_2 B_{a0} B_{r0} \Delta_r^* \Delta_r \label{L1}
\end{align}
where $a_2, c_0, c_2$ are real and $b_2$ could be complex. In the following, we use Eqs.~\eqref{L0} and \eqref{L1} to study homogeneous backgrounds, Gaussian noise, and linear collective modes at tree level. These equations respect the imposed symmetries, but do not include all allowed terms at the displayed orders in derivatives and $a$-fields. In particular, we omit cubic terms involving only $B_{r\mu}$ and $B_{a\mu}$, such as $B_{r0}B_{ai}(B_{ai}+\i\beta\partial_0B_{ri})$ and the other terms required by KMS symmetry, as well as quadratic terms in $B$ with two derivatives, such as $B_{a0}\nabla^2B_{r0}$ and $B_{a0}\partial_0\partial_iB_{ri}$. The cubic terms are allowed in a general medium, but their coefficients vanish at tree level in the probe Abelian holographic model considered here (see Section~\ref{lesson_holo_model}), where the bulk gauge action is quadratic and the coefficients are evaluated about a vanishing scalar background. The quadratic terms with two derivatives need not vanish. Under the critical scaling $\partial_0\sim\nabla^2\sim b_0\sim|\Delta_r|^2$, they give higher-order corrections to the leading soft modes, while they can affect the response at finite frequency and momentum. Appendix~\ref{app:operator_completion} discusses these additional terms and the changes in the background and fluctuation equations when $c_0,c_2$ are nonzero \cite{kamenev_2023,Bu:2022esd}.

The physical meaning of each term in \eqref{L0} and \eqref{L1} is deferred to Section \ref{Stochastic_Equation_eft}, where the dynamical equations are derived to render the meaning of the various terms more transparent.

\section{Stochastic formalism for superconducting phase transition} \label{Stochastic_Equation_eft}

A key advantage of the SK formalism is that it naturally generates the stochastic noise terms required by the FDT, rather than adding them phenomenologically. In this section, we convert the SK EFT of Section \ref{EFT_action} into stochastic equations describing superconducting phase transitions.

In order to illustrate the basic idea, we consider an SK Lagrangian containing Gaussian noises only, which can be written schematically as\footnote{For a complex variable $\Delta_a$, the Gaussian noise part shall be understood as $\Delta_a \Delta_a^*$.}
\begin{align}
\mathcal L = - \Phi_a E[\Phi_r] + \i \lambda \Phi_a \Phi_a \label{L_Phi}
\end{align}
where $E[\Phi_r]$ is a local functional of the physical variable $\Phi_r$. By the Hubbard-Stratonovich transformation, we have an identity \cite{kamenev_2023}
\begin{equation}
\exp\left( - \int d^4x\, \lambda\Phi_a \Phi_a \right) = \mathcal C \int [D\eta] \exp\left\{ - \int d^4x \left( \frac{1}{4\lambda} \eta^2 - \i \Phi_a \eta \right) \right\} \label{HS_transform}
\end{equation}
where $\eta(x)$ is an auxiliary field and the constant $\mathcal C$ determined by the normalization condition
\begin{align}
1 = \mathcal C \int [D\eta] \exp\left\{ - \int d^4x \frac{1}{4\lambda}\left(\eta + 2 \i \lambda \Phi_a\right)^2 \right\}
\end{align}
In the partition function, the identity \eqref{HS_transform} facilitates the trade of the path integral over $\Phi_a$ for the path integral over the auxiliary variable $\eta$,
\begin{align}
Z = & \int [D\Phi_r][D\Phi_a] \exp\left\{ \i \int d^4x \left(- \Phi_a E[\Phi_r] + \i\lambda \Phi_a \Phi_a \right) \right\} \nonumber \\
= & \mathcal C  \int [D\Phi_r][D\eta] \int [D\Phi_a]\exp\left( - \frac{1}{4\lambda} \int d^4x \, \eta^2 \right) \exp\left\{ \i \int d^4x \, \Phi_a \left( -E[\Phi_r] + \eta \right) \right\} \nonumber \\
= & \mathcal C \int [D\eta] \exp\left( - \frac{1}{4\lambda} \int d^4x \, \eta^2 \right) \int [D\Phi_r]\, \delta \left( -E[\Phi_r]+ \eta \right)
\end{align}
The functional delta-function in above expression enforces the following stochastic equation
\begin{align}
	E[\Phi_r] = \eta \label{stochastic_Phi}
\end{align}
with $\eta$ being Gaussian noise, whose distribution is according to the weight
\begin{align}
P_\eta = \exp\left( - \frac{1}{4\lambda} \int d^4x \, \eta^2(x) \right)
\end{align}
The primary conclusion is that the equations of motion obtained from the variation of \eqref{L_Phi} may be reduced to the standard stochastic form \eqref{stochastic_Phi} via the identification $\eta = 2\i \lambda \Phi_a$.

We now apply the idea elaborated above to the specific SK action \eqref{L0} and \eqref{L1}. The variational problem yields the equations of motion
\begin{align}
\frac{\delta S_{sc}}{\delta A_{a\mu}} = 0 \Rightarrow \partial_\nu F_r^{\mu\nu} = J^\mu, \qquad \frac{\delta S_{sc}}{\delta \phi_a} = 0 \Rightarrow \partial_\mu J^\mu = 0, \qquad \frac{\delta S_{sc}}{\delta \mathcal O_a^*} = 0 \Rightarrow J_\Delta = 0 \label{eom_SC}
\end{align}
where
\begin{align}
J^{0}=& a_0B_{r0} + a_2B_{r0}B_{r0} + c_0\Delta_r^*\Delta_r + 2c_2B_{r0} \Delta_r^*\Delta_r, \nonumber\\
J^{i}=& a_1\partial_i B_{r0} - a_1 {\mathcal E}_i - \i b_3 \left[ \left( \mD_i \Delta_r \right)^* \Delta_r - \left(\mD_i \Delta_r \right) \Delta_r^* \right] + \xi^i, \nonumber \\
J_{\Delta}= &b_0\Delta_r +b_1\partial_0\Delta_r +b_2\partial_0^2\Delta_r -b_3 \mD_i\left(\mD_i\Delta_r\right) +c_0B_{r0}\Delta_r +c_1\Delta_r\Delta_r^*\Delta_r +c_2B_{r0}^2\Delta_r + \tilde \zeta \label{Jmu_J_Delta}
\end{align}
where we have identified ($\tilde\zeta$ should be coloured noise)
\begin{align}
\xi^i \equiv - 2 \i \beta^{-1} a_1 B_{ai}, \qquad \tilde \zeta \equiv -2\i\beta^{-1}\mathrm{Re}(b_1)\Delta_a + 2\beta^{-1} {\rm Im}(b_2)\partial_0 \Delta_a
\end{align}
whose distributions are according to the weights
\begin{align}
P_\xi = \exp \left( \int d^4x \frac{\beta}{4a_1} \xi^i(x) \xi^i(x) \right), \qquad P_{\tilde \zeta} = \exp \left( \int d^4x \tilde \zeta^*(x)\frac{\beta}{2 {\rm Re}(b_1) + \i 2{\rm Im}(b_2) \partial_0} \tilde \zeta(x) \right) \label{weight_xi_zeta}
\end{align}
The inclusion of the second-order time derivative for $\Delta_r$ renders the associated stochastic noise colored. This colored-noise kernel $1/(2\re{(b_1+\i 2\im (b_2)\partial_0)})$, implies non-local memory effects in time, characteristic of non-Markovian dynamics.  Standard techniques in stochastic processes allow this system to be mapped to a local Markovian one by introducing an auxiliary Ornstein-Uhlenbeck variable.

Indeed, the static terms in \eqref{eom_SC} and \eqref{Jmu_J_Delta} can be generated via the functional derivatives of a free energy functional 
\begin{align}
\mathcal F = & \int d^3x \left( -b_0 \Delta_r^*\Delta_r - b_3(\mD_i\Delta_r)^*(\mD_i\Delta_r) - \frac{1}{2} c_1\Delta_r^* \Delta_r \Delta_r^* \Delta_r + \frac{\mathcal{B}^2}{2} \right.\nonumber \\
& \qquad \qquad  \left. - \frac{1}{2} a_0 B_{r0}^2 - \frac{1}{3} a_2 B_{r0}^3 - c_0 B_{r0} \Delta_r^*\Delta_r - \frac{1}{2} c_2 B_{r0}^2 \Delta_r^*\Delta_r \right) \nonumber \\
= & \int d^3x \left( -b_0 \mathcal O_r^*\mathcal O_r - b_3 (\bar\mD_i \mathcal O_r)^*(\bar \mD_i \mathcal O_r) - \frac{1}{2} c_1 \mathcal O_r^* \mathcal O_r \mathcal O_r^* \mathcal O_r + \frac{\mathcal{B}^2}{2} \right.\nonumber \\
& \qquad \qquad  \left. - \frac{1}{2} a_0 B_{r0}^2 - \frac{1}{3} a_2 B_{r0}^3 - c_0 B_{r0} \mathcal O_r^* \mathcal O_r - \frac{1}{2} c_2 B_{r0}^2 \mathcal O_r^* \mathcal O_r \right) \label{free_energy_generalized}
\end{align}
where $\mathcal{B}$ is the magnetic field, and the first line inside the parentheses is the GL free energy density. Interestingly, as is clear from the second equality, when the free energy \eqref{free_energy_generalized} is expressed in terms of the variables $A_{r\mu}$ and $\mathcal O_r$, the gauge parameter $\phi_r$ is absent from the GL counterpart, although it will show up in the generalized one. If we had included a term $\mathcal E^2/2$ in the free energy density above, we would generate $- \vec \nabla \cdot \vec {\mathcal E}$ and $\partial_0 \vec{\mathcal E}$ in the Maxwell equations. This shall be thought of as non-stationary and is thus usually omitted in the free energy density.

To ensure that the EFT defines a well-posed initial value problem, we must establish the stability and causality constraints on the Wilsonian coefficients. Firstly, spatial fluctuations must cost positive energy, requiring $b_3 < 0$. A stable superconducting ground state further requires the effective quartic potential to be bounded from below, which requires $c_1-c_0^2/a_0<0$ according to \eqref{background_SC_phase_modified}. Secondly, dissipative (rather than exponentially divergent) stochastic dynamics and non-negative noise spectral densities impose the conditions in \eqref{ImS>0}. When $b_2$ is complex, stability and propagation cannot be inferred from $\mathrm{Re}(b_2)$ alone; they must be checked from the retarded poles of the full fluctuation equations. The holographic coefficients in Section \ref{lesson_holo_model} (see \eqref{holo_coefficients}) obey the elementary sign constraints above and provide a nontrivial consistency check within that model.

This is an appropriate juncture to elaborate on the physical interpretation of each term in \eqref{L0} and \eqref{L1}:

\noindent $\bullet$ From the constitutive relation for $J^0$ in \eqref{Jmu_J_Delta}: $a_0$ is the linear charge susceptibility and the $c_0$-term the superfluid density; the $a_2$-term represents a nonlinear correction to the linear response of charge density to chemical potential (captured by the $a_0$-term) and the $c_2$-term may be interpreted as a superfluid density correction to the linear charge susceptibility. Identifying $B_{r0}$ as the chemical potential $\mu$, we may schematically split $J^0$ as
$J^0 = \chi\left(\mu, |\Delta_r| \right) \mu + J^0_s$, with $\chi$ the generalized charge susceptibility encoding both nonlinear effects and superfluid contributions, and $J^0_s$ stands for the superfluid density ($c_0$-term).

\noindent $\bullet$ From the constitutive relation for $J^i$ in \eqref{Jmu_J_Delta}: $-a_1$ is the electrical conductivity (or equivalently normal charge diffusion constant), while $b_3$-term accounts for the supercurrent contribution to the total U(1) current.

\noindent $\bullet$ From the expression for $J_\Delta$ in \eqref{Jmu_J_Delta}: $b_0$ is identified as the squared mass of the order parameter $\Delta_r$, while $b_1$ characterizes its relaxation time; the $b_2$-term constitutes a second-order time-derivative correction to the dynamics of $\Delta_r$, thereby converting pure diffusive relaxation into wave-like dynamics \cite{Grigorishin:2021gil}; $b_3$-term acts as a kinetic term for $\Delta_r$; the $c_1$-term denotes quartic self-interaction of $\Delta_r$; $c_0$-term and $c_2$-term induce coupling between the scalar gauge potential and order parameter in the evolution equation for $\Delta_r$.

The set of equations \eqref{eom_SC}, armed with \eqref{Jmu_J_Delta} and \eqref{weight_xi_zeta}, determines the temporal and spatial evolutions of the superconducting system near the critical temperature. A virtue of using $B_{r\mu}$ and $\Delta_r$ lies in the manifest demonstration of the gauge invariance. Apparently, a local gauge transformation
\begin{align}
A_{r\mu} \to A_{r\mu} + \partial_\mu \lambda_r, \qquad \phi_r \to \phi_r - \lambda_r, \qquad \mathcal O_r \to e^{\i q \lambda_r} \mathcal O_r
\end{align}
leaves the set of equations \eqref{eom_SC} and \eqref{Jmu_J_Delta} unchanged.

With $\phi_r$ treated as a gauge parameter and specified via a certain gauge convention, \eqref{eom_SC} reduce into a set of equations for the gauge potential $A_{r\mu}$ and order parameter $\mathcal O_r$, as in the usual TDGL framework. Here, as a simple illustrative example, we demonstrate this by neglecting the part $\mathcal L_1$. In a transparent way, the Maxwell equations of \eqref{eom_SC} are,
\begin{align}
\vec \nabla \cdot \vec {\mathcal E} = J^0_{\rm GL}, \qquad \vec \nabla \times \vec {\mathcal B} - \partial_0 \vec {\mathcal E} = \vec J_{\rm GL} + \vec \xi \label{Maxwell_eom_GL}
\end{align}
where $\vec {\mathcal{E} }$ is the electric field. The mean field part of conserved current is
\begin{align}
J^0_{\rm GL} = \chi \mu, \qquad J_{\rm GL}^i = -a_1 (\mathcal E_i - \partial_i \mu) -\i b_3 \left[\mathcal O_r (\bar \mD_i \mathcal O_r)^* - \mathcal O_r^* (\bar \mD_i\mathcal O_r) \right] \label{Jmu_GL}
\end{align}
where $\bar \mD_i \equiv \partial_i - \i q A_{ri}$. Meanwhile, the last equation in \eqref{eom_SC} can be cast into that for the order parameter field $\mathcal O_r$
\begin{align}
b_0\mathcal O_r +b_1(\partial_0 + \i q \partial_0 \phi_r )\mathcal O_r -b_3 \bar\mD_i\left(\bar\mD_i\mathcal O_r\right) +c_1\mathcal O_r \mathcal O_r^* \mathcal O_r + \zeta =0 \label{Or_eom_GL}
\end{align}
where $\zeta = \tilde \zeta e^{-\i q \phi_r} $ obeying the same distribution as specified by $P_{\tilde \zeta}$ (with $b_2$ dropped out)
\begin{align}
P_{\zeta} = \exp \left( \int d^4x \frac{\beta}{2 {\rm Re}(b_1)} \zeta(x) \zeta^*(x) \right)
\end{align}
Interestingly, in the order parameter's evolution equation \eqref{Or_eom_GL}, we see the natural emergence of Gorkov-Eliashberg anomalous term \cite{kamenev_2023}, say a coupling $\partial_0 \phi_r \, \mathcal O_r$.

We now briefly remark on some specific gauge conventions and the corresponding forms of TDGL equations. We will focus on the order parameter's equation.

\noindent $\bullet$ In the gauge convention $\partial_0 \phi_r =0$, the equation \eqref{Or_eom_GL} takes a simple form
\begin{align}
-b_1\partial_0\mathcal O_r = b_0\mathcal O_r -b_3 \bar\mD_i\left(\bar\mD_i\mathcal O_r\right) +c_1\mathcal O_r \mathcal O_r^* \mathcal O_r + \zeta
\end{align}
which is a form widely adopted in the literature. In this gauge choice, we have $B_{r0} \to A_{r0}$.

\noindent $\bullet$ In the gauge convention $\partial_0 \phi_r =- A_{r0}$, the equation \eqref{Or_eom_GL} takes a standard form
\begin{align}
-b_1\bar \mD_0 \mathcal O_r = b_0\mathcal O_r -b_3 \bar\mD_i\left(\bar\mD_i\mathcal O_r\right) +c_1\mathcal O_r \mathcal O_r^* \mathcal O_r + \zeta
\end{align}
where $\bar \mD_0 \equiv \partial_0 -\i q A_{r0}$. This is another commonly employed form in the literature, well motivated by the gauge invariance. In this gauge choice, $B_{r0} \to 0$, which simplifies the constitutive relations for $J^\mu$.

\noindent $\bullet$ The last gauge convention we will discuss is as follows
\begin{align}
\partial_0 J_n^0 + \partial_i J_n^i = 0 \Rightarrow \partial_0 \phi_r - D \partial_i^2 \phi_r = - A_{r0} + D \partial_i A_{ri}
\end{align}
where $J_n^0$ and $J_n^i$ are normal parts of \eqref{Jmu_GL}, and $D = -a_1/a_0$ is the diffusion constant. This is indeed the so-called $\mathcal K$-gauge \cite{kamenev_2023}, which was employed \cite{kamenev_2023} to derive the TDGL framework from the microscopic BCS theory via the integrating out of electronic degrees of freedom.

\section{Collective modes and dispersion relations} \label{collective_modes}

In this section, we study the dynamical modes and analyze the dispersion relations for collective modes based on the EFT.

In the high-temperature phase with $T \gtrsim T_c$, all dynamical fields are treated as small perturbations around zero. Throughout this section, Fourier modes are proportional to $e^{-\i\omega t+\i\vec k\cdot\vec x}$, so stable retarded poles satisfy $\im\omega\leq0$. Linearizing the deterministic parts of \eqref{eom_SC}, we obtain the corresponding dispersion relations
\begin{align}
&\text{electromagnetic wave in metal:} \qquad & &\omega^2 = \i a_1 \omega + k^2; \nonumber \\
&\text{wave-like order parameter $\Delta_r$:} \qquad & &b_2 \omega^2 = - \i b_1 \omega + b_3 k^2 + b_0 \nonumber \\
&\text{wave-like order parameter $\Delta_r^*$:} \qquad & &b_2^* \omega^2 = - \i b_1^* \omega + b_3 k^2 + b_0 \label{EM_wave_Delta_normal_phase}
\end{align}
The sign in the first line is fixed by retarded stability. With $\sigma\equiv-a_1>0$, its two small-momentum poles are
\begin{align}
\omega_{\rm soft}=-\i\frac{k^2}{\sigma}+O(k^4),
\qquad
\omega_{\rm fast}=-\i\sigma+\i\frac{k^2}{\sigma}+O(k^4),
\label{normal_em_poles}
\end{align}
and hence both lie in the lower half-plane.

For the order parameter, define $\delta(k)=b_3k^2+b_0$. The two poles of the second line of \eqref{EM_wave_Delta_normal_phase} are
\begin{align}
\omega_{\pm}(k)=\frac{-\i b_1\pm\sqrt{-b_1^2+4b_2\delta(k)}}{2b_2},
\label{normal_scalar_poles}
\end{align}
while the conjugate equation gives the paired poles $-\omega_{\pm}^*$. For $|\delta|\ll|b_1^2/b_2|$, the two solutions in Eq.~\eqref{normal_scalar_poles} can be expanded as
\begin{align}
\omega_{\rm soft}=-\i\frac{\delta(k)}{b_1}+O(\delta^2),
\qquad
\omega_{\rm fast}=-\i\frac{b_1}{b_2}+\i\frac{\delta(k)}{b_1}+O(\delta^2).
\label{normal_scalar_expansion}
\end{align}
In particular, at $b_0=0$ the strict long-wavelength soft pole is $\omega_{\rm soft}=-\i(b_3/b_1)k^2+O(k^4)$ rather than a mode linear in $k$. A wave-like form can arise when $|4b_2b_3k^2|\gg|b_1^2|$ while the derivative expansion remains valid:
\begin{align}
\omega_{\pm}(k)\simeq-\frac{\i b_1}{2b_2}\pm\sqrt{\frac{b_3}{b_2}}\,k+\cdots.
\label{normal_scalar_inertial}
\end{align}
For real spatial momentum, propagation and damping must be read separately from a pole: $v_g=d\re\omega/dk$ and $\Gamma=-\im\omega$. Thus, if a local dispersion contains a complex coefficient of $k$, its real part gives the wave-packet propagation slope, whereas its imaginary part gives momentum-dependent attenuation or growth. The complex number $\sqrt{b_3/b_2}$ is therefore not, by itself, an observable velocity.

When the system is cooled into the low-temperature phase where $T < T_c$, the above results no longer hold. Indeed, the coefficient $b_0$ turns positive upon $T \lesssim T_c$, indicating that the wave-like dynamical behavior of the order parameter transitions to a regime of linear instability: specifically, for $T \lesssim T_c$, the positive imaginary part of the frequency $\omega$ obtained from the second equation of \eqref{EM_wave_Delta_normal_phase} signals a linear instability that induces the superconducting phase transition. In alternative words, this linear instability means that a new ground state with non-trivial background for the dynamical fields emerges for $T< T_c$, which spontaneously breaks the electromagnetic gauge symmetry.

We now turn our attention to the physics in the low-temperature phase. To streamline the analysis, we adopt two assumptions: 1) neglect higher-order terms by setting $a_2 = c_0 = c_2 =0$; 2) assume that $b_1$ and $b_2$ are real-valued. We will revisit this set of assumptions in subsequent discussions. We consider the static and spatially homogeneous configuration for the new ground state. Then, from \eqref{eom_SC} and \eqref{Jmu_J_Delta}, we have 
\begin{align}
\bar B_{r0} =0, \qquad \bar \Delta_r \equiv \psi_0 = 0~~\text{or} ~~ \pm \sqrt{-\frac{b_0}{c_1}} \label{background_SC_phase}
\end{align}
Apparently, a non-trivial solution with $\psi_0 \neq 0 $ corresponds to the new ground state in the low-temperature superconducting phase. Next, we turn on perturbations on top of the non-trivial background in \eqref{background_SC_phase}
\begin{align}
B_{r\mu} = 0 + \delta B_{r\mu}, \qquad  \Delta_r = (\psi_0 + \delta\psi ) e^{\i q \theta_r}  \label{perturbation}
\end{align}
where $\delta \psi$ is identified as the Higgs mode, and $\theta_r$ is the phase fluctuation of the order parameter. 

Linearizing the imaginary part of order parameter's equation in \eqref{eom_SC} and \eqref{Jmu_J_Delta}, we obtain
\begin{align}
b_1 \partial_0 \theta_r = -b_3 \partial_i ( \delta B_{ri} - \partial_i \theta_r), \label{theta_Cri}
\end{align}
Here we have taken the additional low-frequency limit $|b_2\partial_0^2\theta_r|\ll|b_1\partial_0\theta_r|$. Without this approximation, the phase equation retains its second-order time derivative. This relation helps to combine the phase fluctuation $\theta_r$ with the dynamical gauge field $\delta B_{r\mu}$ through
\begin{align}
	\delta C_{r\mu} \equiv \delta B_{r\mu} - \partial_\mu \theta_r    \label{Crmu_definition}
\end{align}
In this simplified analysis with real $b_1,b_2$, we further assume $\delta C_{r0}=0$. Since $\delta C_{r\mu}$ is gauge invariant, this assumption restricts the fluctuations rather than fixes a gauge. We then have
\begin{align}
	& 0 = \partial_\mu \partial^\mu \delta C_{ri} - \partial_i \partial_j \delta C_{rj}  - \frac{a_1 b_3}{b_1} \partial_i \partial_j \delta C_{rj}  + a_1 \partial_0 \delta C_{ri} + 2b_3 q\psi_0^2 \delta C_{ri}, \nonumber \\
	& 0 = b_2 \partial_0^2 \delta \psi + b_1 \partial_0 \delta \psi - b_3 \partial_i^2 \delta \psi +\left[b_0 + 3c_1 (\psi_0)^2\right]\delta \psi  \label{Cri_psi_eom}
\end{align}
where we have substituted $\partial_0 \theta_r$ with $\partial_i \delta C_{ri}$ using the relation \eqref{theta_Cri}. Upon assuming the same Fourier convention as above, namely $\delta C_{ri}, \delta \psi \sim e^{-\i \omega t + \i \vec k \cdot \vec x}$, the linearized equations \eqref{Cri_psi_eom} yield the dispersion relations of the dynamical modes in the superconducting phase
\begin{align}
	&\text{Transverse photon:} \qquad &&\omega^2 -k^2 -\i a_1 \omega - \frac{2b_0 b_3}{c_1} =0 \nonumber \\
	&\text{Longitudinal photon:} \qquad &&\omega^2 + \frac{a_1 b_3}{b_1} k^2 -\i a_1 \omega - \frac{2b_0 b_3}{c_1} =0 \nonumber \\
	&\text{Higgs mode:} \qquad && \omega^2 - \frac{b_3}{b_2} k^2 + \i  \frac{b_1}{b_2} \omega  + 2 \frac{b_0}{b_2} = 0  \label{EM_Higgs_SC}
\end{align}
The dispersion relations in Eq.~\eqref{EM_Higgs_SC} are obtained for real $b_1,b_2$ and with the additional approximations stated above. They agree with the phenomenological results of \cite{Grigorishin:2021gil,Jeong:2023las} in this limit, with the coefficients determined here by the displayed SK effective action.

Neglecting the medium effect ($a_1=0$) in the first two equations of \eqref{EM_Higgs_SC} recovers the standard Higgs mechanism: the electromagnetic field acquires the Meissner mass $\omega_A^2=2b_0b_3/c_1$. For nonzero $a_1$, the term $-\i a_1\omega$ describes Ohmic damping. In the simplified longitudinal equation, $-a_1b_3/b_1$ is the coefficient of $k^2$, but its square root does not generally give the group velocity as $k\to0$ for a gapped or overdamped mode. The expression derived with real $b_1$ cannot be directly applied to its complex holographic value.

The last equation in \eqref{EM_Higgs_SC} describes the decoupled amplitude mode for real $b_1,b_2$. The coefficient $b_1$ describes relaxation, while $b_2$ gives a second-order time derivative. The quantity $\sqrt{b_3/b_2}$ describes a dispersion slope only when the $k^2$ and $\omega^2$ terms dominate within the range of validity of the EFT. It does not generally give the long-wavelength group velocity of the gapped amplitude mode.

Near the critical regime $T \simeq T_c$, the constant terms in \eqref{EM_Higgs_SC} approach zero as $(T_c - T)$. Then, the medium effect renders the notion of mass gap for both gauge field and Higgs mode subtle \cite{Grigorishin:2021gil,Jeong:2023las}. This issue can be clarified by solving the dispersion equations \eqref{EM_Higgs_SC} at zero wave-vector. The solutions are
\begin{align}
	&\text{photon}: & & \omega = -\frac{1}{2} \i \sigma \pm \sqrt{\omega_A^2-\frac{1}{4} \sigma^2}, & &\qquad \text{with} \quad \sigma \equiv -a_1, \quad \omega_A^2 \equiv \frac{2b_0 b_3}{c_1}, \nonumber \\
	& \text{Higgs mode}: && \omega = -\i \gamma \pm \sqrt{\omega_H^2 - \gamma^2}, & &\qquad \text{with} \quad \gamma \equiv \frac{b_1}{2b_2}, \quad \omega_H^2 \equiv -  2 \frac{b_0}{b_2}.  \label{omega_EM_Higgs}
\end{align}
Without the medium effects (i.e., $\sigma = \gamma =0$), both the gauge field and the Higgs field are massive due to the spontaneously breaking of gauge symmetry. However, the coefficients $\sigma, \gamma$ representing the medium effects are finite near the critical regime. So, near the critical regime, we have $\omega_A << \sigma, \, \omega_H << \gamma$ so that the solutions in \eqref{omega_EM_Higgs} become
\begin{align}
	&\text{photon}: & & \omega \simeq - \i \frac{\omega_A^2}{\sigma}, \qquad \qquad \omega \simeq -\i \sigma, \nonumber \\
	& \text{Higgs mode}: && \omega \simeq -\i \frac{\omega_H^2}{2\gamma}, \qquad \qquad \omega \simeq -2\i \gamma.  \label{omega_EM_Higgs1}
\end{align}
Here, the two solutions for both the photon and Higgs fields become purely imaginary, implying that the strong damping effects near the critical temperature render these modes strongly overdamped, thereby obscuring the observation of a genuine mass gap. To connect these EFT parameters with experimental observables, we note that the superconducting plasma frequency $\omega_A$ is inversely proportional to the London penetration depth, $\lambda_{\rm L}\propto1/|\psi_0|=\sqrt{-c_1/b_0}$ in the simplified sector. According to GL paradigm, the mass parameter $b_0$ scales as $b_0\propto (T_c-T)$ near the critical temperature, and thus the EFT predicts $\omega_A\propto(T_c-T)^{\frac{1}{2}}$, recovering critical divergence of the London penetration depth $\lambda_L\propto(T_c-T)^{-\frac{1}{2}}$. Meanwhile, the parameter $\gamma$ describes the amplitude relaxation rate of the superconducting condensate. It is linked to pair-breaking and recombination processes and may be measured via state-of-the-art ultrafast terahertz pump-probe spectroscopy \cite{PhysRevLett.107.177007,doi:10.1126/science.1254697}.

We now provide a brief discussion of the two assumptions employed above in our analysis to derive the dispersion relations \eqref{omega_EM_Higgs}. To begin with, we consider the effect of higher-order terms by activating the representative coefficient $c_0$ in \eqref{Jmu_J_Delta}. Then, the non-trivial background solution in \eqref{background_SC_phase} gets modified to
\begin{align}
	\bar B_{r0} = \frac{c_0}{a_0} \frac{b_0}{\tilde c_1}, \qquad \bar \Delta_r \equiv \psi_0 =  \pm \sqrt{- \frac{b_0}{\tilde c_1}} \qquad \text{with} \quad \tilde c_1 \equiv c_1 - \frac{c_0^2}{a_0} \label{background_SC_phase_modified}
\end{align}
The $c_0$ coupling also mixes the amplitude and charge fluctuations in the linearized equations. Thus, replacing $c_1$ by $\widetilde c_1$ accounts for the change in the background, but does not describe all changes in the collective modes. The Gauss equation must also be included in the fluctuation analysis. The terms proportional to $c_0$ are essential for recovering the stochastic equations of Model F for critical superfluids, as demonstrated in Appendix~\ref{superfluid_dynamics}.

Next, we consider complex-valued $b_1$ and $b_2$. Their imaginary parts couple the amplitude and phase fluctuations, which in turn mix with the longitudinal gauge field. The modes must therefore be obtained from the coupled linearized equations, including the second-order time derivatives and the Gauss equation. These equations are given in Appendix~\ref{app:complex_modes}. For real momentum, propagation and damping are described separately by $v_g=d\re\omega/dk$ and $\Gamma=-\im\omega$. In particular, a complex coefficient in the dispersion relation should not itself be interpreted as an observable velocity.

When $c_0,c_2$ are nonzero, both the homogeneous background and the coupled fluctuation equations are modified. Appendix~\ref{app:operator_completion} gives these changes and discusses the additional terms required at the same order. A full finite-momentum analysis with the holographic coefficients requires determining the coefficients of these additional terms and is left for future work.

\section{Lesson from a holographic model for superconductor} \label{lesson_holo_model}

In this section, we provide an explicit realization of the EFT construction presented in Section \ref{EFT_action} within a simple holographic superconductor model. We also determine the Wilsonian coefficients in the EFT Lagrangians \eqref{L0} and \eqref{L1} for this model.

In the holographic framework, a gauge symmetry in AdS space corresponds to a global symmetry of the dual field theory residing on the asymptotic boundary of the bulk AdS spacetime. Accordingly, the prototypical holographic setup \cite{Gubser:2008px,Hartnoll:2008kx,Hartnoll:2008vx} (see \cite{Herzog:2009xv,Cai:2015cya} for reviews on this topic), which involves a complex scalar field $\Psi$ coupled to a U(1) gauge field $A_M$ in an asymptotically AdS$_5$ black brane geometry, corresponds to superfluidity in the boundary field theory. Later on, the holographic superfluid model of \cite{Gubser:2008px,Hartnoll:2008kx,Hartnoll:2008vx} was appropriately revised in order to study a superconducting phase transition, see \cite{Domenech:2010nf,Baggioli:2023tlc}. The basic idea \cite{DeWolfe:2020uzb,Ahn:2022azl,Ahn:2025ptl} is to impose a Neumann boundary condition when solving the bulk gauge field $A_M$ so that the boundary gauge field becomes dynamical as required for a superconductor.

\begin{figure}[H]
\centering
\includegraphics[width=1\textwidth]{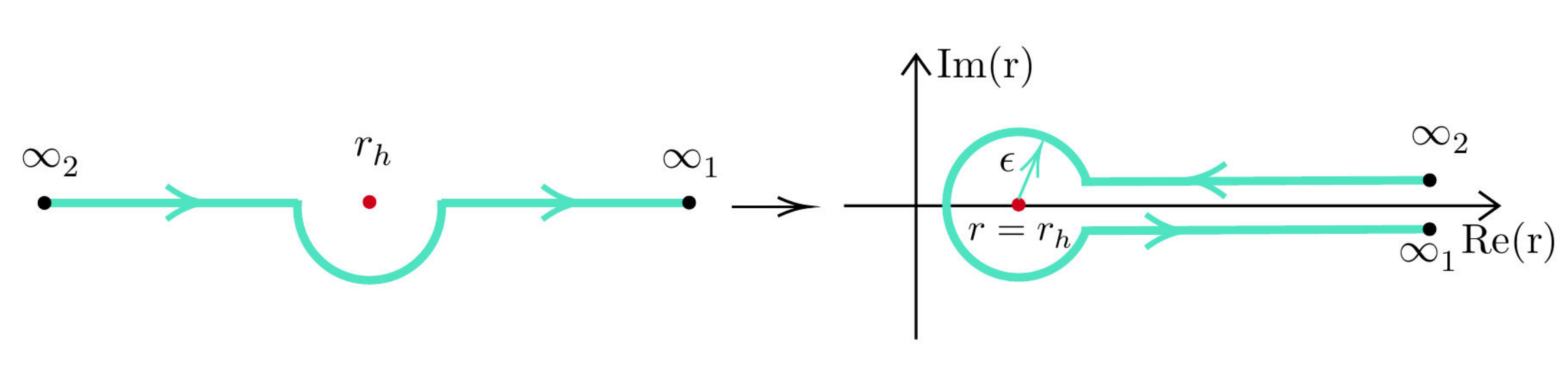}
\caption{Left: complexified double AdS (analytically continued near the horizon) \cite{Crossley:2015tka}; Right: the holographic SK contour \cite{Glorioso:2018mmw}. The two horizontal legs overlap with the real axis.}
\label{holo_contour}
\end{figure}

The total action for a minimal holographic superconductor is
\begin{equation}
S = S_0 + S_{\text{bdy}},  \label{total5Daction}
\end{equation}
where the bulk term $S_0$ is
\begin{equation}
S_0 = \int d^5 x \sqrt{-g} \left[ -\frac{1}{4} F^{MN} F_{MN} - D_M \Psi \left( D^M \Psi \right)^* - m_0^2 \Psi^* \Psi \right]
\label{5Dbulkaction}
\end{equation}
where $D_M = \nabla_M - \i q A_M$, and a $^*$ denotes charge conjugate. To obtain analytical solutions for the bulk fields, we follow \cite{Herzog:2010vz} and adopt the special scalar mass parameter $m_0^2 = -4/L^2$. For convenience,  both the AdS radius $L$ and the coupling constant $q$ will be set to unity in subsequent calculations. The background spacetime is the Schwarzschild-AdS$_5$, with its line element in the ingoing Eddington-Finkelstein coordinate system given by
\begin{align}
	ds^2 = 2dvdr - r^2f(r) dr^2 + r^2 \delta_{ij} dx^i dx^j, 
\end{align}
where the bulk coordinates are denoted as $x^M = (r, x^0, x^i)$ with $v\equiv x^0$ the time coordinate. The blackening factor is given by $f(r) = 1-r_h^4/r^4$ with $r=r_h$ the location of event horizon. The Hawking temperature is given by $T = r_h/\pi$, which is identified with the temperature of boundary field theory. In the context of holographic SK prescription \cite{Liu:2018crr,Glorioso:2018mmw}, the radial coordinate $r$ evolves along the holographic contour depicted in Figure \ref{holo_contour}.

The boundary term $S_{\text{bdy}}$ is:
\begin{align}
	S_{\text{bdy}} = \int d^4 x \sqrt{-\gamma}  &\left\{  -\frac{1}{4} F_{\mu\nu} F^{\mu\nu} \log r + 2\Psi^* \Psi - \frac{\Psi^* \Psi}{\log r} + n_M \left( \Psi^* \nabla^M \Psi + \Psi \nabla^M \Psi^* \right) \right.   \nonumber \\
	& \left. \quad -\frac{1}{4}F_{\mu\nu}F^{\mu\nu} \right\}\Bigg |_{r\to\infty_2}^{r\to\infty_1}  \label{5Dbdyaction}
\end{align}
where $\gamma = \text {det}(\gamma_{\mu\nu})$ with $\gamma_{\mu\nu}$ the induced metric on the boundary hypersurface $\Sigma_s$ located at $r = \infty_s$, and $n_M$ is the outward normal vector of the boundary hypersurface $\Sigma_s$. The boundary term \eqref{5Dbdyaction} plays two crucial roles: it removes the divergences at the asymptotic boundary and guarantees the well-posedness of the bulk variational problem. By the latter, we mean the on-shell variation of bulk action \eqref{total5Daction} takes the desired form \cite{Ahn:2022azl,DeWolfe:2020uzb,Bu:2021clf,Bu:2024oyz}
\begin{align}
	\delta S = \int d^4x \left( \mathcal J^\mu \delta B_\mu + \psi_b^* \delta \Delta + \psi_b \delta \Delta^* \right) \label{delta_S_on-shell}
\end{align}
where by on-shell we have imposed the bulk equations of motion dictated by the bulk action \eqref{5Dbulkaction}. Here, the various quantities in \eqref{delta_S_on-shell} are encoded in the near-boundary asymptotic expansion of the bulk fields
\begin{align}
	A_\mu(r \to \infty, x^\alpha) & = B_\mu(x^\alpha) + \cdots + \frac{\mathcal J_\mu(x^\alpha)}{r^2} + \cdots, \nonumber \\
	\Psi(r\to \infty, x^\alpha) & = \psi_b(x^\alpha)\frac{\log r}{r^2} + \frac{\Delta(x^\alpha)}{r^2} + \cdots, \nonumber \\
	\Psi^*(r\to \infty, x^\alpha) & = \psi_b^*(x^\alpha)\frac{\log r}{r^2} + \frac{\Delta^*(x^\alpha)}{r^2} + \cdots \label{boundary_expansion}
\end{align}
Notice that in both \eqref{delta_S_on-shell} and \eqref{boundary_expansion} we have suppressed the SK indices in all quantities.

Obviously, the precise implementation of a dynamical electromagnetic field on the boundary relies on the on-shell variation of the total bulk action \eqref{total5Daction}, as demonstrated in \eqref{delta_S_on-shell}. By imposing the mixed boundary condition $\Pi^{\mu}-\partial_{\nu}F^{\mu\nu}=0$, where the $\Pi^{\mu}=n_M F^{M\mu}|_{r\to\infty}$ is the boundary surface contribution by bulk action $S_0$, the boundary gauge field ceases to be an external source and becomes a fully fluctuating dynamical degree of freedom \cite{Ahn:2022azl}.

In the saddle point approximation of holographic duality, the partially on-shell bulk action, obtained by substituting the classical solutions to the bulk dynamical equations into \eqref{5Dbulkaction}, is identified with the effective action for the boundary theory \cite{Bu:2021clf,Bu:2024oyz}, namely
\begin{align}
	S_{sc} = S_0\left[A_M[B_\mu,\, \Delta, \, \Delta^*],\, \Psi[B_\mu,\, \Delta,\, \Delta^*] \right] + S_{\rm bdy}
\end{align}
In \cite{Bu:2024oyz}, a perturbative approach was developed to analytically solve the bulk dynamics governed by \eqref{5Dbulkaction}.  Here, we outline the basic idea of \cite{Bu:2024oyz} for solving the bulk dynamical equations. First, create a finite density state in the high-temperature phase, which corresponds to the following static background for bulk fields
\begin{align}
	\bar A_r = - \frac{\bar A_v}{r^2f(r)}, \qquad \bar A_v = \bar \mu \left( 1- \frac{r_h^2}{r^2} \right), \qquad \bar A_i = 0, \qquad \bar \Psi = \bar \Psi^* = 0 \label{homogeneous_background_bulk}
\end{align}
where the first equation corresponds to a gauge choice. Here, $\bar \mu$ has the physical meaning of chemical potential. It was realized that \cite{Herzog:2010vz} only when $q\bar\mu = 2r_h$, can one obtain analytical solutions for bulk perturbation to be introduced later. Thus, throughout this work, we will take
\begin{align}
	\bar \mu = \bar \mu_c + \delta \bar \mu,  \qquad  \text{with}  \qquad  q\bar \mu_c =2r_h
	\end{align}
where $\delta \bar \mu$ represents a perturbation to the critical chemical potential $\bar \mu_c$, which drives the system a little bit away from the critical point. In other words, tuning $\bar \mu$ to $\bar \mu_c$ from below places the system at the critical point for the onset of superconductivity.  Then, turn on general perturbations on top of the background \eqref{homogeneous_background_bulk}
\begin{align}
	A_M(r,x^\alpha) = \bar A_M(r) + \delta A_M(r,x^\alpha), \qquad \qquad  \Psi(r,x^\alpha) = 0+ \delta \Psi(r, x^\alpha)
	\end{align}
where the bulk perturbations $\delta A_M$ and $\delta \Psi$ form a system of nonlinear partial differential equations. In accord with the spirit of EFT, these equations may be solved (semi)-analytically via expansion in terms of the derivative as well as the amplitude of the boundary data $B_\mu, \Delta$. For the computation of most coefficients in \eqref{L0} and \eqref{L1}, it is sufficient to work exactly at the critical point corresponding to $\bar \mu = \bar \mu_c$, yet the coefficient $b_0$ requires a slight deviation from the critical chemical potential $\bar \mu_c$.

Recycling the relevant results from \cite{Bu:2024oyz}, we reproduce the EFT Lagrangians \eqref{L0} and \eqref{L1}, with holographic predictions for various coefficients given by (in unit of $r_h =1$)
\begin{align}
	&a_0=2, \qquad a_1=- 1, \qquad a_2 = 0, \qquad  b_0 = \frac{1}{2} (\bar \mu - \bar \mu_0) \simeq (T_c - T), \nonumber \\
	&  b_1=-\frac{1}{4}(1-3\i), \qquad b_2=  -\frac{1+2\log 2}{8} - \i \frac{\log 2}{8}, \qquad  b_3=-\frac{1}{4}, \nonumber\\
	& c_0=\frac{1}{2}, \qquad c_1= - 0.0208333, \qquad c_2=-0.346573  \label{holo_coefficients}
\end{align}
The holographic result is consistent with the complex-valued $b_1,b_2$ allowed by the EFT. The fact that $a_2 = 0$ may be attributed to the probe limit (i.e., no backreaction of the bulk gravity) undertaken in the holographic study. We expect that $a_2$ will be nonzero beyond probe limit.

\section{Summary and Outlook} \label{sec:summary}

In this work, we formulate a Schwinger-Keldysh effective field theory for an s-wave superconductor near its critical temperature $T_c$. This theory describes the coupled dynamics of a complex scalar order parameter (superconducting condensate) and a dynamical U(1) gauge field. Based entirely on symmetry principles, this novel formalism enables systematic treatment of dissipation and fluctuations, providing an ideal theoretical framework for exploring non-stationary dynamics of superconductors in the critical regime. We extend recent developments of the Schwinger-Keldysh formalism for critical superfluidity \cite{Kapustin:2022iih,Donos:2023ibv,Bu:2024oyz,Donos:2025jxb}, where the gauge field acts only as a background source, and show that gauging critical superfluidity yields critical superconductivity.

A key advantage of the SK formalism is its tractable treatment of real-time dynamics, which is indispensable for studying the non-equilibrium physics of superconductors. In the Gaussian approximation for fluctuation fields, we recast the effective field theory into a stochastic formalism in a manifestly gauge-invariant manner. The resulting stochastic equations describe the spatiotemporal evolution of the superconducting condensate and the dynamical gauge field near the critical regime. Our results clarify several important issues concerning the widely used phenomenological TDGL framework, such as its gauge invariance and the complex nature of the relaxation time. Additionally, based on symmetry principles, we naturally extend the TDGL equations, e.g., by including higher-order nonlinear terms and the second-order time derivative of the order parameter.

Using derived stochastic equations, we analyze the dynamical modes of a critical superconductor. We clarify how the complex-valued relaxation time of the order parameter affects its dynamics, identifying an oscillatory relaxation behavior. We also revisit the spontaneous breaking of electromagnetic gauge symmetry in a non-stationary context: via the Higgs mechanism, phase fluctuation of the order parameter is absorbed into the gauge field. This produces a massive gauge field and a massive Higgs mode (the order parameter’s amplitude) when the system is not too close to the critical point. However, near the critical regime, both the Higgs mode and the gauge field exhibit overdamped diffusive behavior as observed in \cite{Grigorishin:2021gil,Jeong:2023las} from phenomenological extensions of the standard TDGL framework. For complex kinetic coefficients, we derive the coupled linearized equations in the broken phase. The exact normal-phase solution shows that the critical soft pole is quadratic in momentum at long wavelength. When a linear-in-$k$ dispersion is applicable, its complex coefficient describes propagation and attenuation separately. We also discuss the changes in the condensate and fluctuation equations due to nonzero $c_0,c_2$, and the additional terms required by KMS symmetry.

We further provide an explicit realization of our EFT construction by applying the holographic SK technique to a minimal holographic superconductor. Imposing Neumann boundary conditions for the bulk gauge field, we promote the boundary gauge field to a dynamical field and derive the effective action of the boundary superconductor. This construction gives a nontrivial consistency check and yields strongly coupled values of the Wilsonian coefficients for the chosen holographic model. Notably, holography predicts a complex-valued relaxation coefficient $b_1$, suggesting oscillatory relaxation dynamics in strongly coupled superconductors. This feature is distinct from the purely real relaxation coefficient commonly assumed in the GL theory for weakly coupled systems.

The present study may be extended in a number of ways. First, it would be interesting to investigate the effects of fluctuations in the non-stationary regime of a critical superconductor \cite{Larkin2005,Kopnin2001}. Such effects are expected to lead to significant modifications of both thermodynamic and transport properties. Second, the effect of temperature variation can be included by coupling the current framework to the stress tensor sector. This would extend the TDGL to include effects of non-uniform temperature and heat conductivity, allowing for the study of thermal gradients within the superconductor \cite{Kapustin:2022iih}. Third, it would also be interesting to adopt the present framework to investigate vortex dynamics and turbulence in holographic superconductors \cite{Montull:2011im,Dias:2013bwa,Montull:2012fy,Salvio:2012at,Salvio:2013jia,Li:2019swh,Su:2022ysc,Yang:2024vga}, particularly in the non-equilibrium and near-critical regimes where complex physical picture is expected to emerge. Furthermore, it is of great theoretical interest and practical significance to extend the EFT and holographic framework for superfluids and superconductors to systems exhibiting nonrelativistic Schr\"odinger symmetry \cite{Son:2008ye,Adams:2011kb,Wang:2013tv}. Last but not least, we aim to apply the framework developed in this work to investigate theoretical aspects of color superconductivity, which is conjectured to exist in QCD matter at extremely low temperatures and high baryon densities \cite{Alford:2007xm}.

\appendix
\section{Critical superfluidity: from SK EFT to Model F} \label{superfluid_dynamics}

In this appendix, we provide details concerning the derivation of stochastic equations describing critical superfluidity, based on the superfluid EFT. While similar analyses exist in the literature \cite{Kapustin:2022iih, Donos:2023ibv}, our derivation proceeds entirely from the EFT action, without ad-hoc assumptions. Furthermore, employing the superfluid EFT, we derive possible extensions to the stochastic equations for Model F describing critical superfluidity.

As clarified in the main text, a crucial distinction between a critical superconductor and a critical superfluid lies in their dynamical contents. For a critical superfluid, the dynamical degrees of freedom include the conserved charge density and the charged order parameter (superfluid condensate). In the EFT language, the conserved charge density is equivalently described by a neutral scalar field $\phi$, while the superfluid condensate is represented by a complex scalar field $\mathcal O$. The partition function for a superfluid  system is schematically written as \eqref{Z_sf}
\begin{align}
	Z_{sf}[A_{r\mu}, m_r; A_{a\mu}, m_a] = \int [D \phi_r] [D \phi_a] [D \mathcal O_r] [D \mathcal O_a] e^{\i S_{sf}[B_{r\mu}, \Delta_r; B_{a\mu}, \Delta_a]} \label{Z_sf1}
\end{align}
where $A_{r\mu}, A_{a\mu}$ and $m_r, m_a$ are external sources for the conserved density and order parameter, respectively. The superfluid's effective Lagrangian defined by $S_{sf} = \int d^4x \mathcal L_{sf}$ is \cite{Bu:2024oyz} (see also \cite{Donos:2023ibv,Kapustin:2022iih})
\begin{align}
	\mathcal{L}_{sf}=&a_0 B_{a0} B_{r0} + a_1 B_{ai} \partial_0 B_{ri} - \i\beta^{-1} a_1 B_{ai}^2 + a_2 B_{a0}B_{r0}B_{r0}  \nonumber \\
	& + b_0 \Delta_a^* \Delta_r + b_0 \Delta_a \Delta_r^*  + b_1 \Delta_a^* \partial_0 \Delta_r + b_1^* \Delta_a \partial_0 \Delta_r^*   \nonumber \\
	& -2 \i \beta^{-1} \mathrm{Re}(b_1) \Delta_a^*\Delta_a + b_3 \left(\mD_i\Delta_a\right)^* \left(\mD_i\Delta_r\right)  + b_3 \left(\mD_i\Delta_a\right) \left(\mD_i\Delta_r\right)^* \nonumber \\
	& -\mathrm{i} b_3 B_{ai} \left(\mD_i\Delta_r\right)^* \Delta_r + \mathrm{i} b_3 B_{ai} \left(\mD_i\Delta_r\right) \Delta_r^* + c_0 B_{a0} \Delta_r^* \Delta_r + c_0 B_{r0} \Delta_a^* \Delta_r \nonumber \\
	&  + c_0 B_{r0} \Delta_a \Delta_r^* + c_1 \Delta_a \Delta_r^* \Delta_r^* \Delta_r + c_1 \Delta_a^* \Delta_r \Delta_r^* \Delta_r  + c_2 B_{r0}^2 \Delta_a^* \Delta_r \nonumber \\
	&  + c_2 B_{r0}^2 \Delta_a \Delta_r^* + 2 c_2 B_{a0} B_{r0} \Delta_r^* \Delta_r + \Delta_a^* s_r + \Delta_a s_r^* + \Delta_r^*s_a + \Delta_rs_a^* , \label{Lsf}
\end{align}
where $s_{r,a}$ are related to the sources $m_{r,a}$ for the order parameter $\mathcal O_{r,a}$ by
\begin{align}
e^{\i q\phi_r} m_r = s_r,\qquad  e^{\i q\phi_r}m_a = s_a
\end{align}
and $B_{r\mu}, B_{a\mu}, \Delta_r, \Delta_a$ are introduced in \eqref{gauge_invariant}, with $\mD_i$ denoting the spatially covariant derivative compatible with $B_{ri}$ \eqref{gauge-covariant_Di}. Here, we ignored second-order time derivatives of the order parameter (i.e., the $b_2$-terms in \eqref{L1}).

In contrast with \eqref{eom_SC} for a superconductor, the dynamical equations for a superfluid system are given by
\begin{align}
	\frac{\delta S_{sf}}{\delta \phi_a} = 0 \Rightarrow \partial_\mu J^\mu = 0, \qquad \qquad  \frac{\delta S_{sf}}{\delta \mathcal O_a^*} \equiv J_{\mathcal O} = 0 \label{SF_eom}
	\end{align}
Here, the conserved current $J^\mu$ is defined as the functional derivative of superfluid effective action with respect to the external source $A_{a\mu}$
\begin{align}
	J^{0}&\equiv \rho=a_0\mu +a_2\mu^2+c_0\mathcal{O}_r\mathcal{O}_r^*+ 2 c_2 \mu\mathcal{O}_r\mathcal{O}_r^* \nonumber \\
	J^{i}&=-a_1 (\mathcal E_i - \partial_i \mu) - \i b_3\left[\left(\bar \mD_i\mathcal{O}_r\right)^*\mathcal{O}_r-\left(\bar \mD_i\mathcal{O}_r\right)\mathcal{O}_r^*\right] -2\i \beta^{-1} a_1 B_{ai} , \label{Jmu_superfluid}
\end{align}
where $\mu \equiv B_{r0}$ is the dynamical chemical potential. In addition, $J_{\mathcal O}$ is given by
\begin{align}
J_{\mathcal{O}} = & m_r + b_0 \mathcal{O}_r + b_1 (\partial_0 + \i q \partial_0 \phi_r)\mathcal{O}_r  -b_3 \bar \mD_i \left(\bar \mD_i \mathcal{O}_r \right) + c_0 \mu \mathcal{O}_r + c_1 \mathcal{O}_r \mathcal{O}_r^*\mathcal{O}_r \nonumber \\
&+ c_2 \mu^2 \mathcal{O}_r -2 \i \beta^{-1} \mathrm{Re}(b_1) \mathcal O_a \label{J_O_superfluid}
	\end{align}
We advance by inverting the first equation in \eqref{Jmu_superfluid} perturbatively
\begin{align}
	\mu=\frac{\rho}{a_0} -\frac{a_2}{a_0^3}\rho^2 - \frac{c_0}{a_0} \mathcal{O}_r^*\mathcal{O}_r - \frac{2c_2}{a_0^2}\rho\mathcal{O}_r^*\mathcal{O}_r + \cdots, \label{rela_mu_rho}
\end{align}
which is useful in rewriting the dynamical equations \eqref{SF_eom} in terms of the charge density $\rho$ and order parameter $\mathcal O_r$. Eventually, we obtain
\begin{align}
	\partial_0\rho = & a_1 \vec \nabla \cdot \vec{\mathcal E} -\frac{a_1}{a_0}\nabla^2\rho +\frac{c_0a_1}{a_0}\nabla^2\left(\mathcal O_r^* \mathcal O_r\right) + \i b_3\left[ \mathcal{O}_r \left(\bar \mD_i \bar \mD_i \mathcal{O}_r\right)^* - \mathcal{O}_r^* \left( \bar \mD_i \bar \mD_i  \mathcal{O}_r\right) \right] \nonumber\\
	& + \frac{a_1a_2}{a_0^3} \nabla^2\rho^2 + \frac{2a_1 c_2}{a_0^2} \nabla^2 (\rho \mathcal O_r^* \mathcal O_r) + \xi_\rho,   \nonumber  \\
	\bar \mD_0 \mathcal O_r =& - \frac{m_r}{b_1}  -\frac{b_0}{b_1}\mathcal{O}_r+\frac{b_3}{b_1} \bar \mD_i \bar \mD_i \mathcal O_r -\frac{c_0}{b_1a_0}\rho \mathcal O_r -\left(\frac{c_1}{b_1}-\frac{c_0^2}{b_1a_0}\right)\mathcal{O}_r^*\mathcal{O}_r\mathcal{O}_r  \nonumber\\
	& -\left(\frac{c_2}{b_1a_0^{2}}-\frac{c_0a_2}{b_1a_0^3}\right)\rho^2\mathcal O_r + \frac{2c_0 c_2}{a_0^2 b_1} \rho \mathcal O_r^* \mathcal O_r \mathcal O_r   - \frac{\i q}{a_0} \rho\mathcal O_r + \i  q \frac{c_0}{a_0}\mathcal{O}_r^*\mathcal{O}_r\mathcal{O}_r \nonumber \\
	& +\i q \frac{a_2}{a_0^3} \rho^2 \mathcal O_r +\i q \frac{2c_2}{a_0^2} \rho \mathcal O_r^* \mathcal O_r \mathcal O_r +\xi_{\mathcal O}   \label{eom_rho_O}
\end{align}
where $\xi_\rho$ and $\xi_{\mathcal O}$ stand for white Gaussian noises for the conserved density $\rho$ and non-conserved order parameter $\mathcal O_r$, respectively. Setting $a_2 = c_2 = 0$ to neglect higher-order nonlinear terms, we find that the stochastic equations \eqref{eom_rho_O} precisely coincide with those of Model F, which are usually compactly written as
\begin{align}
	\partial_0 \mathcal O_r  & = - 2 \Gamma \frac{\delta \mathcal H}{\delta \mathcal O_r^*} +\i g \mathcal O_r \frac{\delta \mathcal H}{\delta \rho} + \xi_{\mathcal O}, \nonumber \\
	\partial_0 \rho & = \lambda \nabla^2 \left( \frac{\delta \mathcal H}{\delta \rho} \right) - 2 g\, {\rm Im} \left( \mathcal O_r^* \frac{\delta \mathcal H}{\delta \mathcal O_r^*} \right) + \xi_\rho  \label{eom_model_F}
\end{align}
Here, $\mathcal H$ is the free energy functional for a critical superfluid
\begin{align}
	\mathcal H= \int d^3 x & \left[ \frac{1}{2} \tau_0 |\mathcal O_r |^2 + \frac{1}{2} |\bar \mD_i \mathcal O_r|^2 + \tilde u_0 |\mathcal O_r|^4 + \frac{1}{2\chi_0} \rho^2 + \gamma_0 \rho |\mathcal O_r |^2 \right. \nonumber \\
	& \quad \left. - \rho A_{r0}  - m_r \mathcal O_r^* - m_r^* \mathcal O_r \right]  \label{H_model_F}
\end{align}
The relevant Wilsonian coefficients in the EFT Lagrangian \eqref{Lsf} are mapped to those in the Model F \eqref{eom_model_F} and \eqref{H_model_F} via
\begin{align}
	 &\Gamma=\frac{b_3}{b_1},\qquad g=2b_3,\qquad \lambda=2a_1b_3,\qquad  \tau_0=\frac{b_0}{b_3},\nonumber\\
     &\tilde u_0=\frac{c_1a_0-c_0^2}{4a_0b_3},\qquad \chi_0=-2a_0b_3,\qquad \gamma_0=\frac{c_0}{2a_0b_3}. \label{coeffcorespond}
\end{align}
Here, we have switched off the external sources $ A_{r\mu}$  and $m_r$ in \eqref{eom_model_F} when performing coefficient matching.

Our results \eqref{eom_rho_O} generalize the standard stochastic framework \eqref{eom_model_F} for critical superfluid.

\section{Collective modes with complex kinetic coefficients}\label{app:complex_modes}

\subsection{Linearized fluctuation equations}

We first consider the same homogeneous background as in Section~\ref{collective_modes}, with
\begin{align}
a_2=c_0=c_2=0,\qquad q=1,\qquad
\bar B_{r0}=0,
\qquad
\psi_0^2=-\frac{b_0}{c_1}.
\label{app_working_sector}
\end{align}
We allow $b_1,b_2$ to be complex and use the Fourier convention of Section~\ref{collective_modes}. Following Eqs.~\eqref{perturbation} and \eqref{Crmu_definition}, the fluctuations are written as
\begin{align}
 B_{r\mu}=\bar B_{r\mu}+\delta B_{r\mu},
\qquad
\Delta_r=(\psi_0+\delta\psi)e^{\i\theta_r},
\nonumber\\
\delta C_{r\mu}=\delta B_{r\mu}-\partial_\mu\theta_r.
\label{app_fluctuation_variables}
\end{align}
Here $\delta\psi$ and $\theta_r$ are the amplitude and phase fluctuations, respectively. Since $\delta C_{r\mu}$ is gauge invariant, setting $\delta C_{r0}=0$ restricts the fluctuations rather than fixes a gauge. We therefore retain $\delta C_{r0}$ in the following. For $k=|\vec k|\neq0$, we define the longitudinal component as $\delta C_{rL}=\hat k_i\delta C_{ri}$, where $\hat k_i=k_i/k$; the two remaining spatial components are transverse to $\vec k$.

For convenience, write $b_j=b_{jR}+\i b_{jI}$ for $j=1,2$, with $b_{jR}=\re b_j$ and $b_{jI}=\im b_j$, and introduce
\begin{align}
X(\omega)&=-b_{2R}\omega^2-\i b_{1R}\omega,
&Y(\omega)&=-b_{2I}\omega^2-\i b_{1I}\omega,
\nonumber\\
S(k)&=b_3k^2-2b_0,
&\omega_A^2&=-2b_3\psi_0^2=\frac{2b_0b_3}{c_1}.
\label{app_XYS}
\end{align}
The functions $X,Y$ contain the time-derivative terms associated with the real and imaginary parts of $b_1,b_2$. The function $S$ contains the spatial-gradient and amplitude restoring terms, while $\omega_A$ is the gauge-field mass parameter introduced in Section~\ref{collective_modes}. Linearizing the scalar equation and its complex conjugate gives
\begin{align}
[X+S]\delta\psi-\psi_0Y\theta_r&=0,
\nonumber\\
Y\delta\psi+\psi_0X\theta_r+\i\psi_0b_3k\delta C_{rL}&=0.
\label{app_scalar_linear}
\end{align}
Thus the imaginary parts of $b_1,b_2$ couple the amplitude and phase fluctuations. Even when these coefficients are real, the phase equation contains a second-order time derivative through $X$. Neglecting this term requires the low-frequency condition $|b_2\omega|\ll|b_1|$ used in deriving Eq.~\eqref{theta_Cri}.

The transverse gauge-field fluctuations decouple from the scalar fluctuations and obey
\begin{align}
\mathcal D_T(\omega,k)=
\omega^2-k^2-\i a_1\omega-\omega_A^2=0.
\label{app_transverse_det}
\end{align}
With $\sigma=-a_1>0$, their poles are
\begin{align}
\omega_{T,\pm}=-\frac{\i\sigma}{2}
\pm\sqrt{k^2+\omega_A^2-\frac{\sigma^2}{4}}.
\label{app_transverse_poles}
\end{align}
They are underdamped when the square-root argument is positive and overdamped when it is negative.

For the longitudinal fluctuations, define $\delta\boldsymbol\Phi_r=(\delta C_{r0},\delta C_{rL},\theta_r,\delta\psi)^T$. The Gauss equation, the longitudinal Maxwell equation, and Eq.~\eqref{app_scalar_linear} then give
\begin{align}
\mathcal M_L(\omega,k)\delta\boldsymbol\Phi_r=0,
\qquad
\mathcal M_L=
\begin{pmatrix}
k^2+a_0 & k\omega & -\i a_0\omega & 0\\
k\omega & K & a_1k\omega & 0\\
0&0&-\psi_0Y&X+S\\
0&\i\psi_0b_3k&\psi_0X&Y
\end{pmatrix},
\label{app_longitudinal_matrix}
\end{align}
where
\begin{align}
K(\omega)=\omega^2-\i a_1\omega-\omega_A^2.
\end{align}
Here $K$ is the frequency-dependent part of the gauge-field equation. Equation~\eqref{app_longitudinal_matrix} retains all four fluctuation variables and the Gauss constraint. Eliminating $\delta C_{r0}$ using the first row introduces the coefficients
\begin{align}
K_g&=K-\frac{k^2\omega^2}{k^2+a_0},
&L_g&=k\omega\left(a_1+\frac{\i a_0\omega}{k^2+a_0}\right),
\label{app_KgLg}
\end{align}
where $K_g$ multiplies $\delta C_{rL}$ and $L_g$ multiplies $\theta_r$ in the resulting longitudinal Maxwell equation. The longitudinal poles satisfy
\begin{align}
\mathcal D_L(\omega,k)=
K_g\big[X(X+S)+Y^2\big]
-\i b_3kL_g(X+S)=0.
\label{app_longitudinal_det}
\end{align}
Multiplying by $k^2+a_0$ gives a sixth-order polynomial in $\omega$. Its static root $\omega=0$ corresponds to a time-independent phase redefinition with no physical fluctuation. After removing this root, five generically nonzero longitudinal poles remain.

\subsection{Pole structure and propagation}

At $k=0$, Eq.~\eqref{app_longitudinal_det} factorizes as
\begin{align}
K(\omega)\left[X(X-2b_0)+Y^2\right]=0.
\label{app_kzero_factor}
\end{align}
The factor $K$ describes the gauge-field poles, while the second factor describes the coupled amplitude and phase fluctuations. At the critical point, the scalar factor becomes $(X+\i Y)(X-\i Y)$ and gives the paired fast roots
\begin{align}
\omega_{\rm fast}^{(1)}=-\i\frac{b_1}{b_2},
\qquad
\omega_{\rm fast}^{(2)}=-\i\frac{b_1^*}{b_2^*}
=-\left(\omega_{\rm fast}^{(1)}\right)^*.
\label{app_fast_roots}
\end{align}
For small positive $b_0$, the slow amplitude root is
\begin{align}
\omega_{\rm amp}=2\i b_0
\frac{\re b_1}{|b_1|^2}+O(b_0^2),
\label{app_slow_amplitude}
\end{align}
which lies in the lower half-plane when $\re b_1<0$.

For real $b_1,b_2$, $Y=0$ and
\begin{align}
\mathcal D_L=(X+S)\left[K_gX-\i b_3kL_g\right].
\label{app_real_factor}
\end{align}
Taking $X\simeq-\i b_1\omega$, $K_g\simeq K$, and $L_g\simeq a_1k\omega$ recovers the longitudinal equation in Eq.~\eqref{EM_Higgs_SC}. These are the additional low-frequency and Gauss-equation approximations used in the simplified result of Section~\ref{collective_modes}.

When the $\omega^2$ and $k^2$ terms dominate within the range of validity of the EFT, a local form $\omega=\alpha k+\cdots$ gives, from the leading scalar terms,
\begin{align}
|b_2|^2\alpha^4-2b_{2R}b_3\alpha^2+b_3^2=0,
\qquad
\alpha^2=\frac{b_3}{b_2}\quad\text{or}\quad
\alpha^2=\frac{b_3}{b_2^*}.
\label{app_inertial_slopes}
\end{align}
The coefficient $\alpha$ can be complex. For real $k$ and a local dispersion $\omega=\omega_0+\alpha k+\cdots$, the group velocity and damping rate are
\begin{align}
v_g=\frac{d\re\omega}{dk}=\re\alpha,
\qquad
\Gamma=-\im\omega.
\label{app_velocity_definition}
\end{align}
Thus $\re\alpha$ describes the local propagation slope, while $\im\alpha$ gives the momentum dependence of attenuation or growth. A wave-packet interpretation further requires sufficiently weak damping. Equation~\eqref{app_inertial_slopes} does not determine the front velocity, which depends on frequencies beyond the present derivative expansion. At strict long wavelength, the normal-phase critical soft pole in Eq.~\eqref{normal_scalar_expansion} is instead quadratic in $k$, with $v_g(0)=0$. Although the broken-phase poles are modified by the coupling between amplitude, phase, and longitudinal gauge fluctuations, propagation and damping are still defined separately by Eq.~\eqref{app_velocity_definition}.

\subsection{Nonzero \texorpdfstring{$c_0,c_2$}{c0,c2} and additional terms}\label{app:operator_completion}

We now include $c_0,c_2$ and allow $a_2$ to be nonzero. The static homogeneous background has $\bar\Delta_r=\psi_0$, $\bar B_{ri}=0$, and a generally nonzero $\bar B_{r0}$. Choosing $\psi_0$ real, the Gauss and scalar equations obtained from $\mathcal L_0+\mathcal L_1$ give
\begin{align}
a_0\bar B_{r0}+a_2\bar B_{r0}^2+c_0\psi_0^2+2c_2\bar B_{r0}\psi_0^2&=0,
\nonumber\\
b_0+c_0\bar B_{r0}+c_1\psi_0^2+c_2\bar B_{r0}^2&=0.
\label{app_general_background}
\end{align}
For $a_2=c_2=0$, these equations reduce to Eq.~\eqref{background_SC_phase_modified}. Hence the same notation $\psi_0$ is used for the condensate in both backgrounds.

Using the fluctuations in Eq.~\eqref{app_fluctuation_variables}, define
\begin{align}
\chi&=a_0+2a_2\bar B_{r0}+2c_2\psi_0^2,
&d&=c_0+2c_2\bar B_{r0},
\nonumber\\
S_g&=b_3k^2+2c_1\psi_0^2,
&\omega_A^2&=-2b_3\psi_0^2.
\label{app_general_definitions}
\end{align}
Here $\chi$ is the charge susceptibility at fixed condensate, and $2\psi_0d$ is the variation of the charge density with the condensate amplitude. The function $S_g$ replaces $S$ in the amplitude equation, and $\omega_A$ is evaluated on the shifted background. The longitudinal fluctuation matrix obtained from the displayed action is
\begin{align}
\mathcal M_L^{\rm retained}\delta\boldsymbol\Phi_r=0,
\qquad
\mathcal M_L^{\rm retained}=
\begin{pmatrix}
k^2+\chi & k\omega & -\i\chi\omega & 2\psi_0d\\
k\omega & K & a_1k\omega & 0\\
\psi_0d & 0 & -\psi_0(Y+\i d\omega) & X+S_g\\
0 & \i\psi_0b_3k & \psi_0X & Y
\end{pmatrix},
\label{app_general_matrix}
\end{align}
where $X,Y$ have the definitions in Eq.~\eqref{app_XYS} and $K=\omega^2-\i a_1\omega-\omega_A^2$. Setting $a_2=c_0=c_2=0$ gives $\chi=a_0$, $d=0$, and $S_g=S$, and recovers Eq.~\eqref{app_longitudinal_matrix}. Thus nonzero $c_0,c_2$ change the Gauss equation and introduce direct coupling between the amplitude, charge, and phase fluctuations.

As discussed after Eq.~\eqref{L1}, additional terms are allowed at the same orders in derivatives and $a$-fields. To indicate their contributions, we write
\begin{align}
\mathcal L_{\rm EFT}
=\underbrace{\mathcal L_0+\mathcal L_1}_{\mathcal L_{\rm retained}}
+\delta\mathcal L_{B^3}
+\delta\mathcal L_{B^2,\partial^2}
+\delta\mathcal L_{\Delta B,{\rm KMS}}.
\label{app_action_completion}
\end{align}
The three additional terms denote cubic interactions involving only $B$-fields, quadratic $B$-terms with two derivatives, and mixed scalar--gauge terms, respectively. Each class includes the terms required by KMS and chemical-shift symmetry. Representative structures in the first two classes are
\begin{align}
\delta\mathcal L_{B^3}
={}&\lambda_\sigma B_{r0}B_{ai}
 (B_{ai}+\i\beta\partial_0B_{ri})
+\lambda_g B_{r0}B_{ai}\partial_iB_{r0}
+\text{KMS/chemical-shift partners},
\nonumber\\
\delta\mathcal L_{B^2,\partial^2}
={}&\lambda_5B_{a0}\nabla^2B_{r0}
+\lambda_6B_{a0}\partial_0\partial_iB_{ri}
+\lambda_7B_{ai}\partial_0\partial_iB_{r0}
\nonumber\\
&+\lambda_8B_{ai}\partial_0^2B_{ri}
+\lambda_9F_{aij}F_{rij}
+\delta\mathcal L_{aa}.
\label{app_representative_operators}
\end{align}
The $\lambda$'s are additional Wilsonian coefficients, constrained by SK unitarity and KMS symmetry; the displayed structures are not an independent or exhaustive basis. The term $\delta\mathcal L_{aa}$ denotes the noise terms quadratic in $a$-fields required by KMS symmetry at the corresponding order. The mixed terms collected in $\delta\mathcal L_{\Delta B,{\rm KMS}}$ describe additional scalar--gauge interactions and noise correlations; their explicit expressions are not needed for the present analytic results. In the holographic model of Section~\ref{lesson_holo_model}, the pure-$B$ cubic coefficients vanish at tree level because the bulk Maxwell action is quadratic and the coefficients are evaluated about a vanishing scalar background. The quadratic $B$-terms with two derivatives need not vanish, but give higher-order corrections to the leading soft modes under the critical scaling $\partial_0\sim\nabla^2\sim b_0\sim|\Delta_r|^2$.

The effect of any additional term on the background must be determined by varying the action with respect to the $a$-fields before setting them to zero. The derivative and noise terms displayed in Eq.~\eqref{app_representative_operators} do not contribute to the static homogeneous background equations. Their quadratic expansion can, however, modify the fluctuation equations. The resulting pole equation can be written as
\begin{align}
\det\!\left[\mathcal M_L^{\rm retained}
+\delta\mathcal M_L^{B^3}
+\delta\mathcal M_L^{B^2,\partial^2}
+\delta\mathcal M_L^{\Delta B,{\rm KMS}}\right]=0.
\label{app_completed_pole_equation}
\end{align}
Here $\delta\mathcal M_L^{B^3}$, $\delta\mathcal M_L^{B^2,\partial^2}$, and $\delta\mathcal M_L^{\Delta B,{\rm KMS}}$ are the corrections to the $4\times4$ longitudinal matrix obtained by varying and linearizing the corresponding three classes in Eq.~\eqref{app_action_completion}. They are functions of $\omega,k$ and the homogeneous background, with coefficients determined by the additional Wilsonian coefficients. They introduce no new fluctuation variables. Only terms with one $a$-field contribute directly to the tree-level retarded matrix; terms quadratic in $a$-fields determine noise correlations, related to dissipation by KMS symmetry.

Equations~\eqref{app_general_background} and \eqref{app_general_matrix} therefore describe the effects of $c_0,c_2$ within the action displayed in the main text. A full finite-momentum calculation at the stated order also requires the coefficients of the additional matrix terms in Eq.~\eqref{app_completed_pole_equation}. Determining these coefficients and studying the resulting coupled poles provide a separate extension of the present analysis.

\acknowledgments

We would like to thank Matteo Baggioli and Zhiwei Li for helpful discussions. This work was supported by the National Natural Science Foundation of China (NSFC) under the grant No. 12375044 and Doctoral Student Program of the Young S$\&$T Talents Cultivation Project, CAST.

\bibliographystyle{JHEP}
\bibliography{reference}

\end{document}